\documentclass[orivec]{llncs}

\usepackage{amsmath}
\usepackage{amssymb}

\usepackage{amsthm}
\usepackage{enumitem}
\usepackage{cite}
\usepackage{graphicx}
\usepackage{booktabs}
\usepackage{multirow}
\usepackage{xcolor}
\usepackage{siunitx}
\usepackage{tikz}
\usetikzlibrary{fit, arrows.meta, positioning, calc, decorations.pathreplacing}
\usepackage{array}
\usepackage{tabularx}
\usepackage{algorithm}
\usepackage{algpseudocode}
\usepackage{xspace}
\usepackage[anycase]{thmtools}

\usepackage{hyperref}
\usepackage[capitalise]{cleveref}

\newcommand{\nbdash}{\nobreakdash-\hspace{0pt}}

\usepackage{orcidlink}
\renewcommand{\orcidID}[1]{\orcidlink{#1}}
\newcommand{\Letter}{{\fontfamily{mvs}\fontencoding{U}\fontseries{m}\fontshape{n}\selectfont\char66}}

\renewcommand{\epsilon}{\varepsilon}
\renewcommand{\phi}{\varphi}
\renewcommand{\vec}[1]{\mathbf{#1}}

\newcommand{\bigO}{\mathcal{O}}
\newcommand{\naturals}{\mathbb{N}}
\newcommand{\reals}{\mathbb{R}}
\newcommand{\posreals}{\reals_{\geq 0}}
\newcommand{\interval}[2]{[#1,#2]}
\newcommand{\setnocond}[1]{\{#1\}}
\newcommand{\setcond}[2]{\{\, #1 \mid #2 \,\}}
\newcommand{\abs}[1]{\lvert #1 \rvert}

\DeclareMathOperator{\range}{range}

\newcommand{\expectation}{\mathbb{E}}
\newcommand{\probability}{\mathbb{P}}

\newcommand{\AP}{\mathit{AP}}
\newcommand{\modelsymbol}[1]{\mathcal{#1}}

\newcommand{\mdp}[1][M]{\modelsymbol{#1}}
\newcommand{\pmdp}[1][M]{\mdp[#1]_{\parametersSet}}

\newcommand{\mstates}{S}
\newcommand{\minit}[1][s]{\bar{#1}}
\newcommand{\mactions}{\mathit{Act}}
\newcommand{\mtransitions}{\mathbf{P}}
\newcommand{\mlabelling}{L}

\newcommand{\parameter}[1][v]{#1}
\newcommand{\parameters}[1][\parameter]{\vec{#1}}
\newcommand{\parametersSet}{\mathrm{V}}
\newcommand{\parametersDomain}{\mathcal{V}}
\newcommand{\evaluation}{\nu}
\newcommand{\evaluate}[2][\evaluation]{#2\langle#1\rangle}

\newcommand{\lP}[2][\opt]{\mathtt{P}^{#1}_{#2}}
\newcommand{\lER}[2][\opt]{\mathtt{R}^{#1}_{#2}}
\newcommand{\lF}{\mathbf{F}}

\newcommand{\lBU}[1]{\mathbin{\mathbf{U}^{\leq #1}}}

\newcommand{\errorRate}{\epsilon}
\newcommand{\significanceLevel}{\eta}
\newcommand{\margin}{\lambda}
\newcommand{\nsamples}{\ell}
\newcommand{\coefficients}[1][c]{\vec{#1}}

\newcommand{\nobjs}{M}
\newcommand{\obj}[1]{\psi_{#1}}
\newcommand{\valuefun}[1]{f_{#1}}
\newcommand{\approxfun}[1]{\tilde{f}_{#1}}
\newcommand{\gapfun}[1]{g_{#1}}
\newcommand{\approxgap}[1]{\tilde{g}_{#1}}
\newcommand{\unsafeSet}[1]{U_{#1}}
\newcommand{\vulnRegion}[1]{V_{#1}}
\newcommand{\vulnMeasure}[1]{q_{#1}}
\newcommand{\severity}[1]{w_{#1}}
\newcommand{\residualRisk}[1][k]{R_{#1}}
\newcommand{\targetRisk}{\epsilon_{\mathrm{target}}}
\newcommand{\priorDist}{W}
\newcommand{\Indicator}{\mathbb{I}}
\newcommand{\discover}[1]{D_{#1}}
\newcommand{\confirmSet}[1]{A_{#1}}
\newcommand{\confirmRisk}[1][k]{Z_{#1}}
\newcommand{\sampleSeq}{\nu_{1},\nu_{2},\ldots}

\newcommand{\pacpma}{\textsc{PacPMA}\xspace}
\newcommand{\storm}{\textsc{Storm}\xspace}
\newcommand{\stormpars}{\textsc{Storm-pars}\xspace}
\newcommand{\prism}{\textsc{PRISM}\xspace}

\crefname{table}{Tab.}{Tabs.}
\Crefname{table}{Table}{Tables}
\crefname{section}{Sect.}{Sects.}
\Crefname{section}{Section}{Sections}
\crefname{theorem}{Thm.}{Thms.}
\Crefname{theorem}{Theorem}{Theorems}
\crefname{lemma}{Lem.}{Lemmas}
\Crefname{lemma}{Lemma}{Lemmas}
\crefname{corollary}{Cor.}{Cors.}
\Crefname{corollary}{Corollary}{Corollaries}
\crefname{proposition}{Prop.}{Pros.}
\Crefname{proposition}{Proposition}{Propositions}
\crefname{definition}{Def.}{Defs.}
\Crefname{definition}{Definition}{Definitions}
\crefname{remark}{Rem.}{Remarks}
\Crefname{remark}{Remark}{Remarks}
\crefname{algorithm}{Alg.}{Algs.}
\Crefname{algorithm}{Algorithm}{Algorithms}

\begin{document}

\title{Scenario-Based Risk Certificates for Multi-Property Vulnerability Discovery in Parametric MDPs}
\titlerunning{Risk Certificates for Parametric MDPs}

\author{
     Zhiming Chi\inst{1,2,3}\orcidID{0009-0003-1110-5203}
     \and
     Andrea Turrini\inst{1,2}\orcidID{0000-0003-4343-9323}
     \and
     Yong Li\inst{1,2}\orcidID{0000-0002-7301-9234}
     \and
     Lijun Zhang\inst{1,2,3}\textsuperscript{(\Letter)}\orcidID{0000-0002-3692-2088}
 }
\institute{
    Key Laboratory of System Software (Chinese Academy of Sciences), Beijing, China
    \and
    Institute of Software, Chinese Academy of Sciences, Beijing, China
    \and
    University of Chinese Academy of Sciences, Beijing, China
    \\
    \email{zhanglj@ios.ac.cn}
}

\maketitle

% ============================================================
\begin{abstract}
Safety-critical systems such as network protocols and distributed algorithms depend on environmental parameters whose exact values are uncertain.
Parametric Markov decision processes (pMDPs) model such systems by making transition probabilities symbolic functions of the parameters.
For a single quantitative requirement, parametric verification asks which parameter valuations satisfy the requirement.
In practice, however, engineers often monitor several safety and performance requirements at once, and one valuation may make several requirements vulnerable simultaneously.
In this paper we study \emph{multi-property vulnerability discovery under requirement-wise semantics}:
given a pMDP and $\nobjs$ quantitative requirements, we construct candidate regions of the parameter space that may contain violations, and study budgets for random testing under a specified parameter distribution.
To solve this, we adopt a two-layered approach in which the scenario-approximation layer provides PAC guarantees on identifying conservative vulnerability regions for each requirement while the testing layer treats random parameter exploration as coverage of overlapping regions and yields a stopping certificate computed from the marginal region masses alone;
a composition theorem then bounds the expected non-discovery score by the scenario approximation error plus the testing risk from the two layers.
For confirmed discoveries, candidate hits are checked by model checking and budgets use independently certified discovery probabilities. Experiments evaluate these certificates using empirical surrogates and independent confidence bounds.
\end{abstract}

% ============================================================
\section{Introduction}
\label{sec:introduction}

Markov decision processes (MDPs) are the standard model for systems that exhibit both probabilistic and nondeterministic behavior, such as network protocols, distributed algorithms, and safety-critical controllers~\cite{DBLP:books/wi/Puterman94}.
When transition probabilities depend on uncertain or configurable parameters, MDPs can be replaced by \emph{parametric} MDPs (pMDPs), in which each parameter valuation induces a concrete MDP~\cite{DBLP:conf/ictac/Daws04,DBLP:journals/sttt/HahnHZ11}.
Given a quantitative requirement $\obj{}$, a central verification task for pMDPs is to determine what parts of the parameter space correspond to MDPs satisfying $\obj{}$ and what parts to MDPs violating $\obj{}$;
we call the latter the \emph{vulnerability region} for $\obj{}$.
When working with a real system, however, one is usually interested in checking it against several safety and performance requirements, to find whether some of them are violated;
the same parameter valuation may expose more than one weakness.

Our question begins once several vulnerability regions must be tested:
if a tester repeatedly draws independent parameter valuations from a specified distribution and checks the requirements, how many samples are needed before the expected number of requirements whose violations remain undiscovered is small?
The number $k$ of fresh, independent valuations is the \emph{random-testing budget}.
A test identifies candidate requirements whose candidate regions contain the sampled valuation; exact model checking can then confirm which requirements are actually violated.
Thus overlapping regions are helpful: the same test can discover several requirements.
Geometrically, one can color the vulnerability region of each requirement with a different color.
A random test is a point in this colored parameter space; every color covering that point is discovered.
For example, a protocol tester may draw packet-loss and interference parameters from a distribution chosen to represent expected deployment environments, and check delivery reliability and delay at each sampled configuration. The distribution describes variation between configurations; the transition probabilities of each instantiated MDP describe random behavior within that configuration. 
Such a distribution may be estimated from operational data or specified as a testing assumption; our experiments use a uniform distribution, not an empirically measured deployment distribution.

The first layer already supplies symbolic candidate regions. If the sole objective is to find one counterexample per requirement, one may instead search these regions directly and confirm candidate points by model checking; random testing is not necessary for that task. Our testing layer addresses a different question: given a policy of independent random testing under a specified distribution, what budget is needed for a desired discovery level? Computing this budget and executing the tests are separate activities, and obtaining the required probability estimates itself has a cost.
The practical stopping question is therefore:
\begin{center}
\fcolorbox{gray}{gray!10}{\parbox{0.85\linewidth}{\centering\smallskip
\emph{After how many random tests $k$ does the expected number of undiscovered requirements fall below a user-specified threshold?}
\smallskip}}
\end{center}

\begin{figure}[tb]
\centering
\begin{tikzpicture}[
    box/.style={draw, rounded corners=3pt, minimum height=0.7cm, minimum width=1.5cm, align=center, font=\scriptsize},
    arr/.style={->, >=stealth, semithick},
    lbl/.style={font=\tiny, text=gray!50!black},
]

\path[use as bounding box] (-1.6,-0.5) rectangle (9.51,2.345);

% ==================== pMDP input ====================
\node[box, fill=gray!10] (pmdp) at (-0.5, 0.85) {pMDP $\pmdp$\\$\nobjs$ requirements};

% ==================== Scenario-approximation layer ====================
\begin{scope}[shift={(1.6, 0)}]
    \coordinate (sem-in) at (0, 0.85);
    \draw[->, semithick] (0,0) -- (2.2,0) node[right, font=\tiny] {$\theta$};
    \draw[->, semithick] (0,0) -- (0,1.7) node[above, font=\tiny] {$\valuefun{j}$};
    \fill[blue!8, opacity=0.6] plot[smooth] coordinates {(0.12,0.48) (0.35,0.62) (0.6,0.82) (0.9,1.07) (1.2,1.42) (1.48,1.65) (1.75,1.55) (2.0,1.12)}
        -- plot[smooth] coordinates {(2.0,0.95) (1.75,1.38) (1.48,1.48) (1.2,1.25) (0.9,0.9) (0.6,0.65) (0.35,0.45) (0.12,0.3)}
        -- cycle;
    \draw[blue!70, thick, smooth] plot coordinates {(0.12,0.3) (0.35,0.45) (0.6,0.65) (0.9,0.9) (1.2,1.25) (1.48,1.48) (1.75,1.38) (2.0,0.95)};
    \draw[red!70, thick, decorate, decoration={brace, amplitude=2.5pt, mirror}]
        (0.97,-0.05) -- (1.985,-0.05) node[midway, below=3pt, font=\tiny, text=red!70!black] {$\vulnRegion{j}$};
    \foreach \x/\y in {0.22/0.35, 0.5/0.55, 0.78/0.8, 1.06/1.05, 1.34/1.4, 1.62/1.5, 1.9/1.25} {
        \fill[blue!70] (\x, \y) circle (1.2pt);
    }
    \node[font=\small\bfseries, text=blue!60!black] at (1.0, 2.2) {Scenario-approximation layer};
    \coordinate (sem-out) at (2.2, 0.85);

    \draw[dashed, gray] (0,1.15) -- (2.0,1.15) node[right, font=\tiny, text=gray] {$\tau_j$};
    % Vertical dotted lines at threshold crossings
    \draw[densely dotted, red!60, thick] (0.97,1.164) -- (0.97,0);
    \draw[densely dotted, red!60, thick] (1.985,1.164) -- (1.985,0);
\end{scope}

% ==================== Conservative regions box ====================
\node[box, fill=blue!8, minimum width=1.5cm, font=\scriptsize] (vr) at (5.3, 0.85) {Conservative\\regions\\$\vulnRegion{1}\cdots\vulnRegion{\nobjs}$};

% ==================== Testing Layer ====================
\begin{scope}[shift={(7.0, 0.15)}]
    \coordinate (test-in) at (0, 0.85);
    \fill[gray!5] (-0.15,-0.15) rectangle (2.5,1.75);
    \draw[gray!30] (-0.15,-0.15) rectangle (2.5,1.75);

    \begin{scope}[rounded corners=4pt]
        \clip (0.08,0.08) -- (0.9,-0.0) -- (1.25,0.35) -- (1.1,1.0) -- (0.35,1.1) -- cycle;
        \fill[fill=red!50!blue, nearly transparent] (0.75,-0.0) -- (1.6,0.08) -- (2.0,0.6) -- (1.7,1.25) -- (0.95,1.2) -- (0.6,0.6) -- cycle;
    \end{scope}

    \draw[draw=blue!40, thick, rounded corners=4pt]
        (0.08,0.08) -- (0.9,-0.0) -- (1.25,0.35) -- (1.1,1.0) -- (0.35,1.1) -- cycle;
    \node[font=\tiny, text=blue!60!black] at (0.45, 0.48) {$\vulnRegion{1}$};

    \draw[draw=red!40, thick, rounded corners=4pt]
        (0.75,-0.0) -- (1.6,0.08) -- (2.0,0.6) -- (1.7,1.25) -- (0.95,1.2) -- (0.6,0.6) -- cycle;
    \node[font=\tiny, text=red!60!black] at (1.5, 0.75) {$\vulnRegion{2}$};

    \foreach \x/\y in {0.42/0.7, 1.3/0.25, 1.85/0.35, 0.92/0.75, 0.15/1.35, 1.5/1.4, 2.2/1.1} {
        \fill[brown] (\x, \y) circle (1pt);
    }
    \fill[green!50!black] (0.42, 0.7) circle (1.5pt);
    \fill[green!50!black] (1.3, 0.25) circle (1.5pt);
    \fill[green!50!black] (0.92, 0.75) circle (1.5pt);

    \fill[brown] (0.05, 1.65) circle (1pt);
    \node[anchor=west, font=\tiny] at (0.05, 1.65) {sample};
    \fill[green!50!black] (1.15, 1.65) circle (1.5pt);
    \node[anchor=west, font=\tiny] at (1.15, 1.65) {discovered};

    \node[font=\small\bfseries, text=green!45!black] at (1.15, 2.05) {Testing Layer};
\end{scope}

% ==================== Arrows (drawn last for correct layering) ====================
\draw[arr] (pmdp.east) -- (1.6, 0.85) node[midway, above, lbl] {query};
\draw[arr] (3.8, 0.85) -- (vr.west);
\draw[arr] (vr.east) -- (6.85, 0.85);

\end{tikzpicture}
\caption{Framework overview.
\textcolor{blue!60!black}{Scenario-approximation layer}: 
the model checker is queried at sampled parameters (dots), and a polynomial proxy (curve) with a one-sided PAC margin (shaded) yields a conservative vulnerability region $\vulnRegion{j}$.
\textcolor{green!45!black}{Testing layer}: 
random draws test the parameter space; region hits are candidates for confirmation, and an overlap (purple) allows one draw to expose several requirements.
}
\label{fig:framework}
\end{figure}
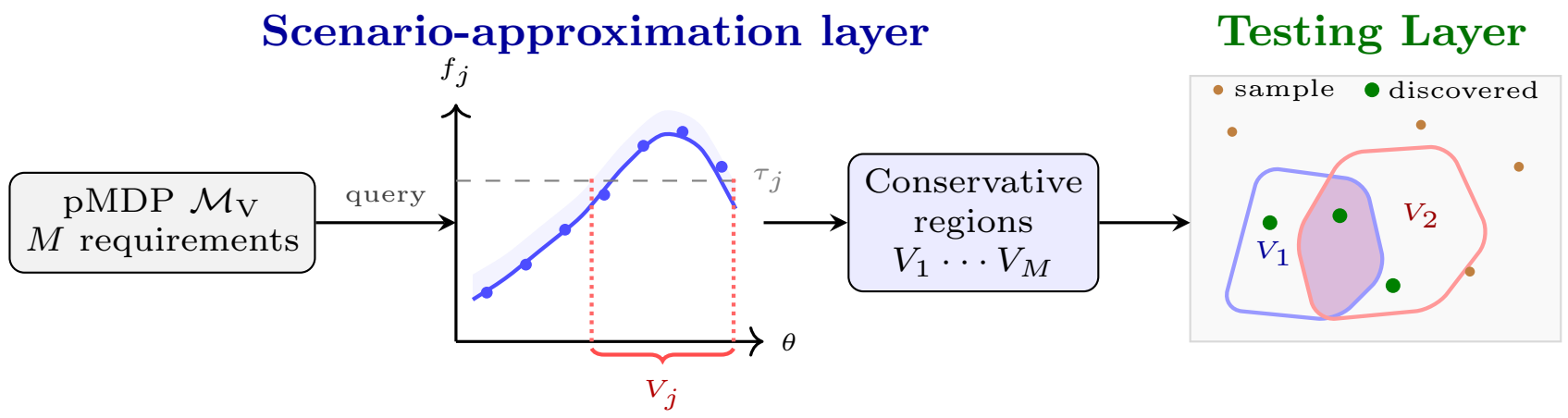

Our solution has two layers, shown in \cref{fig:framework}.
The first layer, the \emph{scenario-approximation layer}, queries the model checker at sampled parameter valuations and fits, for each requirement, a low-degree polynomial proxy for its value function.
A one-sided PAC margin turns this proxy into a conservative region that is cheap to evaluate and is designed to cover all but a small prior mass of true violations.
The second layer, the \emph{testing layer}, uses the masses and overlaps of these regions to quantify how many random draws are needed to discover the vulnerable requirements.
There are then two distinct sources of non-discovery:
the approximation may omit a small part of a true violating region, or the random tests may fail to hit the conservative region that represents it;
we bound their combined effect by our composition theorem.

The vulnerability region of each of the $\nobjs$ requirements is represented by a polynomial whose degree $d$ is chosen by the user.
The testing layer turns these regions into a coupon-collector problem with overlapping coupon types.
We instantiate classical inclusion--exclusion to obtain the exact expected discovery time and derive a \emph{marginal-mass} stopping certificate that is evaluated in time $\bigO(\nobjs)$ from the marginal region masses alone.
The composition theorem~\ref{thm:endToEnd} then combines the PAC approximation error with the residual risk after $k$ random~tests.

The decomposition rests on a semantic choice we call \emph{requirement-wise semantics}:
each value function is computed under its own per-requirement extremal scheduler, i.e., the entity resolving nondeterminism in the pMDP, rather than requiring a single scheduler to optimize all requirements jointly.
This is different from the common scheduler used in multi-objective synthesis.
For the vulnerability-discovery question we ask, requirement-wise semantics captures a useful worst-case view:
a system is vulnerable if there is \emph{some} scheduler that makes \emph{at least one} requirement fail.
It suffices that each requirement can be violated under its own worst-case scheduler;
we do not need a single scheduler that simultaneously violates all requirements.
This observation is what makes the parametric setting manageable: each requirement can be analyzed independently with existing tools, and the results composed at the testing layer.
The two layers address different quantities: scenario coverage controls omitted violation mass, while testing risk describes discoveries under the specified sampling distribution.
The result is an end-to-end bound with a single confidence parameter.

The contribution of our paper is threefold:
\begin{enumerate}
\item
  We formalize multi-property vulnerability discovery under requirement semantics, prove a \emph{coverage guarantee} (\cref{prop:coverage}), and establish an explicit \emph{composition theorem} (\cref{thm:endToEnd}) that combines the scenario approximation error with the expected testing risk under a single confidence parameter.
\item 
    We instantiate classical inclusion-exclusion for overlapping vulnerability regions to obtain an exact expected discovery time, and we derive a \emph{marginal-mass} stopping certificate that bounds the expected number of undiscovered requirements using only the marginal region masses (\cref{sec:risk}).
\item 
    We evaluate our solution on ten benchmark families with up to five monitored requirements, and we measure controlled model-size scaling across all ten families (\cref{sec:benchmarks}).
    After the candidate regions are fixed, independent exact model checking samples provide simultaneous bounds on omitted violations and confirmed-discovery probabilities; separate repetitions of random parameter testing evaluate the resulting stopping certificate.
\end{enumerate}

\paragraph*{Organization of the paper.}
\cref{sec:related-work} positions our contribution against related work.
\cref{sec:preliminaries} recalls pMDPs, the coupon collector problem, and the scenario approach.
\cref{sec:framework} introduces the signed-gap formalization and the certified vulnerability regions.
\cref{sec:risk} develops the overlap-aware testing theory and the certification algorithm.
\cref{sec:benchmarks} presents the benchmark suite, the experimental results, and a comparison against alternative testing strategies.
\cref{sec:conclusion} concludes the paper.
All missing proofs are provided in Appendix~\ref{app:missingProofs}.

% ============================================================
\section{Related Work}
\label{sec:related-work}

\begin{table}[tb]
\caption{Comparison of our approach with the related work.}
\label{tab:related}
\centering
\normalsize
\setlength{\tabcolsep}{3.5pt}
\renewcommand{\arraystretch}{1.08}
\begin{tabularx}{\linewidth}{@{}>{\raggedright\arraybackslash}m{2.75cm}>{\raggedright\arraybackslash}m{3.15cm}>{\raggedright\arraybackslash}m{2.25cm}>{\raggedright\arraybackslash}X@{}}
\toprule
Method / Tool & Model and semantics & Parameter-region output & Guarantee \\
\midrule
\prism{}~\cite{DBLP:conf/cav/KwiatkowskaNP11};\allowbreak \storm{}~\cite{DBLP:journals/sttt/HenselJKQV22} & MDP; common scheduler & none (concrete model) & none \\
\hline
PROPhESY~\cite{DBLP:conf/cav/DehnertJJCVBKA15};\allowbreak \stormpars{}~\cite{DBLP:journals/fmsd/JungesAHJKQV24} & pMC/pMDP; separate requirements & exact partition per requirement & none \\
\hline
DiPS~\cite{HajnalSP21} & pMC; joint data constraints over PCTL requirements & adaptive box partition & data-derived CIs \\
\hline
PacPMA~\cite{DBLP:conf/atva/LiuTHXZ23} & pMC; one requirement & approximate region & per-requirement PAC \\
\hline
Badings et al.~\cite{DBLP:journals/sttt/BadingsCJJKT22} & pMDP; one requirement & approximate region & per-requirement PAC \\
\hline
Forejt et al.~\cite{DBLP:conf/atva/ForejtKP12}; Hartmanns et al.~\cite{Hartmanns2020MultiCost} & MDP; common scheduler & Pareto/multi-cost tradeoff & none \\
\midrule
\textbf{This paper} & pMDP; independent requirements & conservative regions & testing + composition \\
\bottomrule
\end{tabularx}
\end{table}

Existing approaches fall into three groups: verification of parametric Markov models, common-scheduler multi-objective verification, and random testing.
A method-by-method summary is given in \cref{tab:related}.

\paragraph*{Parametric Markov models.}
Several exact methods construct parameter regions for individual requirements.
Examples are PARAM~\cite{DBLP:journals/sttt/HahnHZ11}, PROPhESY~\cite{DBLP:conf/cav/DehnertJJCVBKA15}, and \stormpars~\cite{DBLP:journals/fmsd/JungesAHJKQV24}.
Related methods include those of Badings et al.~\cite{DBLP:conf/cav/BadingsJMTJ23}, Heck et al.~\cite{DBLP:conf/atva/HeckQSKJ25}, and Cubuktepe et al.~\cite{DBLP:journals/tac/CubuktepeJJKT22}; Jansen et al.~\cite{DBLP:conf/birthday/0001JK22} provide a survey.
When these methods complete, they give stronger region information and need no random testing to establish that region.
The scenario approach~\cite{DBLP:journals/tac/CalafioreC06,DBLP:journals/arc/CampiGP09,CampiGaratti2018WaitJudge} instead fits an approximation from sampled valuations.
Liu et al.~\cite{DBLP:conf/atva/LiuTHXZ23} provide PAC polynomial approximations for pDTMCs, Chi et al.~\cite{Chi2025} extend this approach to pMDPs, and Badings et al.~\cite{DBLP:journals/sttt/BadingsCJJKT22} study scenario-based verification of uncertain pMDPs.
These methods certify one approximation at a time; the present work additionally gives one testing budget for discovering several possibly overlapping regions.

DiPS~\cite{HajnalSP21} combines observations with several PCTL requirements for a pMC.
It recursively divides the parameter domain into boxes and classifies each box using confidence intervals derived from the observations.
Our setting differs in two direct ways: the model may contain nondeterminism, and the requested output includes a bound on how many fresh parameter tests are needed to discover the individual vulnerability regions.
Robust MDPs~\cite{NilimElGhaoui2005} and bounded-parameter MDPs~\cite{GivanLeachDean2000} optimize against sets of transition probabilities, while statistical model checking~\cite{SenViswanathanAgha2004,YounesSimmons2002,DBLP:series/lncs/LegayLTYSG19} samples executions of one fixed model rather than valuations of uncertain parameters.

\paragraph*{Common-scheduler multi-objective verification.}
Multi-objective verification asks whether one scheduler can satisfy several objectives jointly~\cite{Etessami2008multiobjective,DBLP:conf/tacas/ForejtKNPQ11,DBLP:conf/atva/ForejtKP12,ChatterjeeMM07}.
Tools such as \prism~\cite{DBLP:conf/cav/KwiatkowskaNP11}, \storm~\cite{DBLP:journals/sttt/HenselJKQV22}, and MultiGain~\cite{DBLP:conf/tacas/BrazdilCFK15} compute Pareto curves or bounded tradeoffs for concrete MDPs; related extensions cover compositional~\cite{DBLP:journals/iandc/KwiatkowskaNPQ13}, continuous-time~\cite{DBLP:conf/euromicro/AlasmariC22}, interval~\cite{DBLP:journals/tomacs/HahnHHLT19}, and parameterized-transition~\cite{DBLP:conf/valuetools/Scheftelowitsch17} settings.
Here the question is whether any requirement can be violated, so each requirement is analyzed under its own worst-case scheduler rather than under one scheduler shared by all requirements.

\paragraph*{Random testing.}
Neal studied the generalized coupon-collector problem~\cite{ccp_generalisedCouponCollectorProblem}, and Shioda derived upper and lower collection-time bounds~\cite{CCP_bounds}; 
both start from specified outcome probabilities.
Arcuri et al.~\cite{ccp_randomtesting} used coupon-collector arguments to analyze the effectiveness and predictability of random testing.
These works are not concerned with probabilistic model checking; 
indeed, they neither construct property-induced vulnerability regions for parametric Markov models nor account for error introduced when those regions are approximated.
Still, we use their results in our testing layer, after model checking and scenario approximation have determined the property-induced vulnerability regions and their probability masses, to accommodate overlapping pMDP vulnerability regions, derive a weighted stopping certificate that needs only their marginal masses, 
and compose its testing risk with the scenario approximation miss mass in an end-to-end guarantee.

% ============================================================
\section{Preliminaries}
\label{sec:preliminaries}

\subsection{Parametric Markov Decision Processes}
\label{ssec:pmdp}

\begin{definition}
\label{def:mdp}
    An MDP is a tuple $\mdp = (\mstates,\minit,\mactions,\mtransitions,\mlabelling)$ where $\mstates$ is a finite nonempty state set, $\minit \in \mstates$ is the initial state, $\mactions$ is a finite action set, $\mtransitions \colon \mstates \times \mactions \times \mstates \to \posreals$ gives transition probabilities, and $\mlabelling \colon \mstates \to 2^{\AP}$ is a labelling function. An action $a$ is enabled at $s$ if $\sum_{s'\in\mstates}\mtransitions(s,a,s')>0$; its outgoing probabilities are the probabilities $\mtransitions(s,a,s')$ of transitions from $s$ to successor states $s'$. For every enabled state-action pair these probabilities sum to one, and every state has at least one enabled action.
\end{definition}
Nondeterminism in MDPs is resolved by a scheduler, inducing a Markov chain and well defined probability distributions over paths~\cite{DBLP:books/daglib/0020348}.

A pMDP replaces numeric probabilities by rational functions over a finite parameter set $\parametersSet = \{\parameter_{1}, \dots, \parameter_{n}\}$.
Intuitively, the parameters represent quantities we do not know exactly (e.g., a failure rate or a network delay).
Each parameter $\parameter$ ranges over a compact interval $\range(\parameter) = \interval{L_{\parameter}}{U_{\parameter}}$, and the parameter domain is $\parametersDomain = \prod_{\parameter \in \parametersSet} \range(\parameter)$.
Every valuation $\evaluation \in \parametersDomain$ induces a concrete MDP $\evaluate{\mdp}$ by instantiating the transition functions~\cite{DBLP:conf/ictac/Daws04,DBLP:journals/sttt/HahnHZ11}.

We consider requirements $\obj{j}$ given as quantitative formulas in PRCTL, such as
$\lP[\max]{=?}[\lF\; \texttt{err}]$ and $\lER[\min]{=?}[\lF\; \texttt{goal}]$;
Here $\lP[\max]{=?}[\lF\;\texttt{err}]$ queries the maximal probability of eventually reaching an error state, and $\lER[\min]{=?}[\lF\;\texttt{goal}]$ queries the minimal expected accumulated reward until reaching a goal. A requirement compares the resulting value with a threshold. The probability and reward operators quantify over paths of the instantiated MDP, with the indicated extremum over schedulers; see~\cite{DBLP:books/daglib/0020348} for the full logic.
For every requirement $\obj{j}$, we write $\valuefun{j} \colon \parametersDomain \to \reals$ for the resulting value function over parameter valuations.
For pMDPs this function is typically piecewise-defined because the optimizing scheduler may depend on the valuation~\cite{DBLP:journals/jcss/JungesK0W21}.
In other words, different parameter regions may have different optimal schedulers, which makes the value function hard to represent analytically when the number of parameters grows.

\subsection{Coupon Collector Problem}
\label{ssec:coupon}

The form of the coupon collector problem (CCP) relevant here is the \emph{generalized} or \emph{non-uniform} CCP, where different coupon types may have different draw probabilities.
It arises naturally in random testing, where each test ``draws'' a parameter valuation and the goal is to discover all target outcomes.
The analogy is as follows: each vulnerability region is a ``coupon type'', each random parameter test is a ``draw'', and we want to know how many draws are needed to collect all coupons, i.e., to discover all vulnerability regions.

\begin{definition}[Generalized coupon collector]
\label{def:coupon}
    Let $C = \setnocond{C_{1}, \dots, C_{\nobjs}}$ be a set of $\nobjs$ coupon types occurring with respective probabilities $p = (p_{1}, \dots, p_{\nobjs})$ such that $p_{j} > 0$ for each $j$ and $\sum_{j=1}^{\nobjs} p_{j} \leq 1$.
    At each step $k$, coupon type $C_{j}$ is drawn with probability $p_{j}$, and the remaining probability $1 - \sum_{1 \leq j \leq M} p_{j}$ corresponds to a null draw that collects no target coupon.
    For the probability vector $p$ above, the \emph{collection time} $T(p)$ is the random minimum step at which every $C_{j}$ has been drawn at least once:
    \[
        T(p) = \inf\setcond{k \in \naturals}{\text{each $C_{j} \in C$ has been drawn in one of the first $k$ steps}}.
    \]
    \end{definition}

The key quantity for statistical guarantees is the \emph{tail distribution}, i.e., the probability that after $k$ draws not all coupon types have been collected.
Let $P_{J} := \sum_{j \in J} p_{j}$ for $J \subseteq \{1, \dots, \nobjs\}$.
The standard inclusion--exclusion formula for the generalized coupon collector problem~\cite{ccp_generalisedCouponCollectorProblem} gives, for every $k>0$,
\begin{equation}
\label{eq:coupon-tail}
    \probability(T(p) > k)
    = \sum_{i=1}^{\nobjs} (-1)^{i+1}
    \sum_{J \subseteq \{1,\dots,\nobjs\}: \abs{J}=i} (1-P_{J})^{k} .
    \end{equation}
The inner sum ranges over all subsets of size $i$, i.e., over the nonempty power-set levels.
Intuitively, $P_{J}$ is the probability that a single draw hits at least one coupon in $J$.
So $(1-P_{J})^{k}$ is the probability that none of those coupons appears in the first $k$ draws.
Inclusion--exclusion then aggregates these terms into the exact tail probability.
Applying the tail-sum identity to \cref{eq:coupon-tail} yields the exact expected collection time for the non-uniform coupon collector problem;
for the uniform case $p_{j} = 1/\nobjs$, this reduces to the familiar harmonic-number formula $\nobjs H_\nobjs = \nobjs \sum_{j=1}^{\nobjs} 1/j$.

In \cref{sec:risk} we reuse exactly this tail-distribution argument, but adapt it to overlapping vulnerability regions (defined in \cref{sec:framework}).
The key difference from the classical CCP is that a single parameter sample may discover several requirements simultaneously.
So the disjoint-case subset sum $P_{J} = \sum_{j \in J} p_{j}$ is replaced by the union mass
\begin{equation}
\label{eq:unionMass}
    Q_{J} = \int_{\bigcup_{j \in J} \vulnRegion{j}} \priorDist(\evaluation)\, d\evaluation,
\end{equation}
where $\vulnRegion{j}$ is the conservative vulnerability region for requirement $\obj{j}$ and $\priorDist$ is the parameter prior (cf.~\cref{def:prior,def:conservative-gap}).
This yields the overlap\nbdash{}aware exact discovery\nbdash{}time formula used in the testing layer (cf.~\cref{thm:generalizedCoupon}).
The precise formula and its proof are given in \cref{thm:generalizedCoupon}.

\subsection{Scenario Approximation}
\label{ssec:scenario}

The scenario approach~\cite{DBLP:journals/tac/CalafioreC06,DBLP:journals/arc/CampiGP09} fits a template function to finitely many sampled valuations and certifies that large approximation errors can occur only on a small fraction of the parameter space.
The idea is simple: rather than trying to compute the rational value function for requirement $\obj{j}$ exactly (which is intractable in general), we evaluate it at a modest number of parameter points, fit a polynomial, and quantify how bad the fit can be.

\begin{definition}[Parameter prior]
\label{def:prior}
    A \emph{parameter prior} is a probability density $\priorDist \colon \parametersDomain \to \posreals$ over the parameter domain $\parametersDomain$.
\end{definition}
We impose no restriction on the shape of $\priorDist$: 
it may be uniform, truncated Gaussian, Beta, or any other distribution supported on $\parametersDomain$.
The density can be evaluated or sampled to estimate the masses of the candidate regions introduced in \cref{sec:framework}, using numerical integration or Monte Carlo sampling.

\begin{remark}
    When $\priorDist$ is the uniform density on $\parametersDomain$, all region masses reduce to normalized Lebesgue volumes.
    The non-uniform case assigns higher weight to operationally more likely valuations, which is the natural semantics in robustness verification.
\end{remark}

\begin{definition}
\label{def:pac-approx}
    Given $\valuefun{} \colon \parametersDomain \to \reals$ and a probability density $\priorDist$ on $\parametersDomain$, a \emph{template function} $\approxfun{}$ of degree $d$ is a polynomial
    $\approxfun{}(\evaluation) = \coefficients^{\mathsf{T}} \phi(\evaluation)$,
    where $\phi(\evaluation) = (1, \parameter_{1}, \dots, \parameter_{n}, \parameter_{1}^{2}, \parameter_{1}\parameter_{2}, \dots)$ denotes the monomial basis of total degree $d$ (comprising $m_{d} = \binom{n+d}{d}$ terms) and $\coefficients \in \reals^{m_{d}}$ collects the polynomial coefficients.
    We call $\approxfun{}$ an \emph{$(\errorRate, \significanceLevel, \margin)$-PAC approximation of $f$} if, with confidence at least $1 - \significanceLevel$,
    \[
        \int_{\setcond{\evaluation \in \parametersDomain}{\abs{\approxfun{}(\evaluation) - \valuefun{}(\evaluation)} > \margin}} \priorDist(\evaluation)\, d\evaluation \leq \errorRate.
    \]
\end{definition}

To fit such a template function from data, let $\setnocond{\evaluation_{i}}_{i=1}^{\nsamples} \subseteq \parametersDomain$ be $\nsamples$ independent parameter samples drawn from $\parametersDomain$ according to $\priorDist$.
Given requirement $\obj{j}$, the polynomial surrogate $\approxfun{j}$ and the margin $\margin_{j}$ are obtained by solving the following scenario optimization problem:
\begin{equation}
\label{eq:scenario-lp}
\begin{split}
\min_{\coefficients_{j},\margin_{j}} \quad & \margin_{j} \\
\mathrm{s.t.} \quad & -\margin_{j} \leq \valuefun{j}(\evaluation_{i}) - \coefficients_{j}^{\mathsf{T}} \, \phi(\evaluation_{i}) \leq \margin_{j}, \quad \forall\, i \in \{1, \dots, \nsamples\}, \\
& \margin_{j} \geq 0,
\end{split}
\end{equation}
where $\coefficients_{j} \in \reals^{m_{d}}$ collects the polynomial coefficients for $\obj{j}$.
The requirement values $\valuefun{j}(\evaluation_{i})$ at each sampled parameter point are obtained by instantiating the pMDP and solving the resulting concrete MDP via a call to a model checker such as \storm~\cite{DBLP:journals/sttt/HenselJKQV22}.
The standard scenario theorem yields the following sample bound for this optimization problem.

\begin{theorem}[\kern-1.1ex\cite{DBLP:journals/arc/CampiGP09}]
\label{thm:scenario-samples}
    Assume the scenario optimization problem~\eqref{eq:scenario-lp} is feasible and has a unique optimum value $\margin$.
    If the number of sampled valuations satisfies
    \[
        \nsamples \geq \frac{2}{\errorRate}\Big(\ln \frac{1}{\significanceLevel} + m_{d} + 1\Big),
    \]
    then the fitted polynomial $\approxfun{j}$ is an $(\errorRate, \significanceLevel, \margin)$-PAC approximation of the target $\valuefun{j}$.
\end{theorem}

In our framework we support two concrete backends for the scenario approximation:
a PAC-based solution or an empirical surrogate.
The PAC-based solution, based on a \pacpma-style backend~\cite{DBLP:conf/atva/LiuTHXZ23}, solves the LP~\eqref{eq:scenario-lp} exactly and obtains $\margin_{j}$ as the optimal objective value, providing a PAC-certified margin.
However, at the sample sizes prescribed by \cref{thm:scenario-samples}, the unregularized LP can overfit the training data, producing polynomials with large coefficients and poor generalization (see Appendix~\ref{app:backend-detail} for more details).
In our experiments we therefore use an empirical surrogate obtained from a \emph{regularized} scenario program:
\begin{equation}
\label{eq:scenario-qp}
\begin{split}
\min_{\coefficients_{j},\margin_{j}} \quad & \margin_{j} + \alpha \lVert \coefficients_{j} \rVert_{2}^{2} \\
\mathrm{s.t.} \quad & -\margin_{j} \leq \valuefun{j}(\evaluation_{i}) - \coefficients_{j}^{\mathsf{T}} \, \phi(\evaluation_{i}) \leq \margin_{j}, \quad \forall\, i \in \{1, \dots, \nsamples\}, \\
& \margin_{j} \geq 0,
\end{split}
\end{equation}
where $\coefficients_{j} \in \reals^{m_{d}}$ collects the polynomial coefficients for $\obj{j}$ and $\alpha > 0$ is a small regularization constant.
The program~\eqref{eq:scenario-qp} is a convex quadratic program (QP):
the penalty $\alpha \lVert \coefficients_{j} \rVert_{2}^{2}$ makes the objective strictly convex, which keeps the fitted coefficients bounded and prevents the overfitting observed for the unregularized LP.
The surrogate is $\approxfun{j}(\parameters) = \coefficients_{j}^{\mathsf{T}} \phi(\parameters)$, and the margin $\margin_{j}$ from the optimum of~\eqref{eq:scenario-qp} equals the maximum training residual.
This empirical margin upper-bounds the worst-case training error but is not itself PAC-certified;
see Appendix~\ref{app:backend-detail} for a detailed comparison.
Under the standing assumptions of \cref{thm:scenario-samples}, either backend yields an $(\errorRate_{j}, \significanceLevel_{j}, \margin_{j})$-PAC approximation; the testing-layer theory in \cref{sec:risk} is independent of the chosen backend.

% ============================================================
\section{Certified Vulnerability Regions}
\label{sec:framework}

In this section we build the scenario-approximation layer of our framework.
The key idea is to convert each scenario approximation into a conservative vulnerability region in the parameter space so that no true violation within the PAC margin is missed.
Informally, we want to paint a ``safe map'' of the parameter space:
every valuation flagged as safe is truly safe (up to the PAC error), and the map is deliberately over-cautious so that real violations are never missed.

We first introduce a signed-gap formulation that unifies upper- and lower-bounded requirements (\cref{ssec:signed-gap}) and then define the conservative regions and prove the requirement-wise coverage guarantee (\cref{ssec:conservative-vuln}).

\subsection{Signed Violation Functions}
\label{ssec:signed-gap}

Let $\obj{1}, \dots, \obj{\nobjs}$ be $\nobjs$ quantitative requirements.
Requirement $\obj{j}$ has threshold $\tau_{j}$ and value function $\valuefun{j}$.
It is either upper-bounded, $\valuefun{j}(\evaluation) \leq \tau_{j}$, or lower-bounded, $\valuefun{j}(\evaluation) \geq \tau_{j}$.

\begin{definition}[Signed violation function]
\label{def:signed-gap}
    For each $j \in \{1, \dots, \nobjs\}$, let
    \[
    \gapfun{j}(\evaluation)=
    \begin{cases}
    \valuefun{j}(\evaluation)-\tau_{j}, & \text{if requirement $\obj{j}$ is upper-bounded,}\\
    \tau_{j}-\valuefun{j}(\evaluation), & \text{if requirement $\obj{j}$ is lower-bounded.}
    \end{cases}
    \]
    The valuation $\evaluation$ violates requirement $\obj{j}$ if and only if $\gapfun{j}(\evaluation)>0$.
    We write
    $\unsafeSet{j} = \setcond{\evaluation \in \parametersDomain}{\gapfun{j}(\evaluation)>0}$
    for the true violating set of requirement $\obj{j}$.
\end{definition}

This signed-gap view lets us treat reward and probability requirements uniformly.
For example, an upper-bounded reachability risk and a lower-bounded success probability both become \emph{positive-gap} events when the requirement is violated.
The advantage is that we only need to reason about the sign of a single function, regardless of the original comparison direction of the requirement.

\subsection{Conservative Vulnerability Regions}
\label{ssec:conservative-vuln}

Assume we have an $(\errorRate_{j},\significanceLevel_{j},\margin_{j})$-PAC approximation $\approxfun{j}$ of $\valuefun{j}$.
We convert it into a conservative signed surrogate by pushing the approximation in the unsafe direction.
The construction is deliberately one-sided: we would rather flag a safe valuation as potentially unsafe than miss a truly unsafe one.

\begin{definition}[Conservative signed surrogate]
\label{def:conservative-gap}
    For each requirement $\obj{j}$, let
    \[
        \approxgap{j}(\evaluation) =
        \begin{cases}
        \approxfun{j}(\evaluation) + \margin_{j} - \tau_{j}, & \text{if requirement $\obj{j}$ is upper-bounded,}\\
        \tau_{j} - \approxfun{j}(\evaluation) + \margin_{j}, & \text{if requirement $\obj{j}$ is lower-bounded.}
        \end{cases}
    \]
    The corresponding conservative vulnerability region and its prior mass are
    $\vulnRegion{j} = \setcond{\evaluation \in \parametersDomain}{\approxgap{j}(\evaluation) > 0}$ and $\vulnMeasure{j} = \int_{\vulnRegion{j}} \priorDist(\evaluation)\, d\evaluation$, respectively.
\end{definition}
The polynomial predicate defining $\vulnRegion{j}$ is the interface between the two layers of the framework.
It allows the testing layer to check whether fresh parameter samples lie in the candidate region without calling the model checker. A candidate hit still requires model checking to confirm a violation, and this check can miss violations outside the candidate region. Thus
the following proposition bounds the true violating mass omitted by this inexpensive predicate.

\begin{restatable}[Requirement-wise coverage]{proposition}{propCoverage}
\label{prop:coverage}
    Given requirement $\obj{j}$, let $\approxfun{j}$ be an $(\errorRate_{j}, \significanceLevel_{j}, \margin_{j})$-PAC approximation of $\valuefun{j}$.
    Then, with confidence at least $1 - \significanceLevel_{j}$,
    \[
        \int_{\unsafeSet{j} \setminus \vulnRegion{j}} \priorDist(\evaluation)\, d\evaluation \leq \errorRate_{j}.
    \]
\end{restatable}

\begin{corollary}[Simultaneous certification]
\label{cor:simultaneous}
    If each requirement $\obj{j}$ uses parameters $(\errorRate_{j}, \significanceLevel_{j})$ and the region $\vulnRegion{j}$ is constructed as in \cref{def:conservative-gap}, then with confidence at least
    $1 - \sum_{1 \leq j \leq \nobjs} \significanceLevel_{j}$,
    all requirement-wise miss-mass bounds in \cref{prop:coverage} hold simultaneously.
\end{corollary}

\begin{remark}
\label{rem:conservative-overlap}
The regions $\vulnRegion{1},\dots,\vulnRegion{\nobjs}$ are not required to be disjoint.
Indeed, having them overlapping is expected: a single parameter valuation may violate several requirements at the same time.
We directly take advantage of such an overlap in the next section, rather than discarding it.
\end{remark}

% ============================================================
\section{Overlap-Aware Testing Risk}
\label{sec:risk}

In this section we develop the testing layer of our framework.
Whereas \cref{sec:framework} provided conservative vulnerability regions in parameter space, we now ask: 
\emph{how many random parameter samples are needed to discover all such regions with high confidence?}
For candidate-region hits, the answer depends on the region masses $q_{j}$ and, for the exact formula, on their overlap structure. Confirmed violations require the corresponding confirmed-event probabilities instead.
Intuitively, regions with large prior mass are more likely to be hit, while small-mass regions require more samples; overlap helps because a single sample can discover several regions at once.
We first derive the exact expected discovery time via inclusion-exclusion (\cref{ssec:exact-risk}), then introduce a weighted non-discovery score that admits a certificate and a stopping rule based only on the marginal region masses (\cref{ssec:residual-risk}), and finally compose both layers into a single composite bound (\cref{ssec:e2e}).

We analyze random testing at the level of the conservative vulnerability regions.
Given the vulnerability regions $\vulnRegion{j}$ for $1 \leq j \leq \nobjs$, let $\sampleSeq$ be i.i.d.\@ draws from the prior density $\priorDist$.
After $k$ samples, the set of requirements whose candidate regions have been hit is
$\discover{k}=\setcond{j \in \{1,\dots,\nobjs\}}{\text{$\exists t \leq k$ such that $\nu_{t} \in \vulnRegion{j}$}}$.

\subsection{Exact Discovery Time}
\label{ssec:exact-risk}

The following result instantiates, for overlapping vulnerability regions, the tail-distribution argument from \cref{eq:coupon-tail}.
While the inclusion-exclusion derivation itself is classical, the instantiation to overlapping scenario-approximated regions and the resulting risk certificate, including the marginal-mass bound that follows, are new.
The key insight is that the tail probability $\probability(T>k)$ is a union of undiscovered-region events.
Inclusion-exclusion converts this union into a sum of geometric series parameterized by the union masses $Q_{J}$ defined in \cref{eq:unionMass}.
Intuitively, $(-1)^{|J|+1}/Q_{J}$ is the expected waiting time for hitting the union of regions indexed by $J$, and the alternating signs correct for over-counting.

Let $T$ denote the first time at which every conservative region has been hit at least once, i.e., 
$T = \inf\setcond{k \in \naturals}{\discover{k} = \setnocond{1, \dots, \nobjs}}$.

\begin{restatable}[Exact expected discovery time]{theorem}{thmGeneralizedCoupon}
\label{thm:generalizedCoupon}
Assuming $Q_{J} > 0$ for every nonempty $J \subseteq \{1,\dots,\nobjs\}$, we obtain
\[
    \expectation[T] = \sum_{\emptyset \neq J \subseteq \{1,\dots,\nobjs\}} (-1)^{|J|+1} \frac{1}{Q_{J}}.
\]
\end{restatable}
This result is the overlapping-region analogue of the non-uniform CCP expectation identity obtained from \cref{eq:coupon-tail}, with the disjoint-case subset sums $P_{J}$ replaced by the union masses $Q_{J}$.

\subsection{Marginal-Mass Undiscovered-Requirement Certificate}
\label{ssec:residual-risk}

For testing, exact full discovery is often too strict.
We therefore introduce a weighted score that only tracks which conservative regions have not yet been exposed---intuitively, ``how many requirements are still invisible to the tester after $k$ draws?''
The score is simple: each undiscovered requirement contributes its severity weight.
Because each region is missed by all $k$ independent draws with probability $(1-q_{j})^{k}$, linearity of expectation yields a closed-form certificate that avoids the inclusion-exclusion sum entirely.
This has a practical payoff: 
the tester does not need to know how regions overlap, only how large each region is.

\begin{definition}[Non-discovery score (residual risk)]
\label{def:residual-risk}
    Let $\severity{j} \in (0,1]$ be the severity weight of requirement $\obj{j}$ and $\Indicator$ be the indicator function.
    The \emph{non-discovery score} (or, \emph{residual risk}) after $k$ samples is defined as
    \[
        \residualRisk = \sum_{j=1}^{\nobjs} \severity{j} \Indicator\big[j \notin \discover{k}\big].
    \]
\end{definition}
Equivalently, requirement $\obj{j}$ contributes weight $\severity{j}$ if and only if none of the first $k$ tests has hit $\vulnRegion{j}$.
Thus $\residualRisk$ counts the weighted number of undiscovered \emph{requirements}, not the prior mass of $\bigcup_{j \notin \discover{k}} \vulnRegion{j}$.
The weights let the user prioritize requirements: a safety-critical requirement may be assigned $\severity{j}=1$, while a performance requirement receives a lower weight, so that the score reflects the relative importance of undiscovered requirements.

\begin{restatable}[Marginal-mass non-discovery certificate]{theorem}{thmResidualRisk}
\label{thm:residualRisk}
    For every testing budget $k \in \naturals$, the expected value of $\residualRisk$ satisfies
    \[
        \expectation[\residualRisk] = \sum_{j=1}^{\nobjs} \severity{j} (1 - \vulnMeasure{j})^{k}
    \]
    and it can be computed in time $\bigO(\nobjs)$ once the region masses $\vulnMeasure{j}$ are known.
\end{restatable}

\begin{corollary}[Stopping rule]
\label{cor:stopping}
    Let $k^{\star} = \min\setcond{k \in \naturals}{\expectation[\residualRisk] \leq \targetRisk}$.
    Then stopping after $k^{\star}$ samples guarantees that the expected weighted number of requirements whose candidate regions remain unhit is at most $\targetRisk$.
\end{corollary}

\begin{remark}
\label{rem:what-is-certified}
    The certificate in \cref{thm:residualRisk,cor:stopping} is stated for the conservative regions $\vulnRegion{j}$.
    The value $\expectation[\residualRisk]$ gives the expected \emph{weighted count} of requirements whose conservative regions have not been hit yet, i.e., the expected number of requirements that remain invisible to the tester after $k$ draws.
    It is \emph{not} the prior mass of the union of still-undiscovered parameter valuations; 
    the two quantities differ because $\residualRisk$ counts requirements, not parameter volumes.
    By \cref{prop:coverage}, the scenario approximation layer contributes an additional requirement-wise miss mass of at most $\errorRate_{j}$ with confidence $1 - \significanceLevel_{j}$;
    the composition theorem (\cref{thm:endToEnd} below) adds this to the testing risk.
    The framework thus deliberately separates two sources of conservativeness:
    the scenario approximation error, and the residual testing risk over the certified regions.
\end{remark}

\begin{remark}[Exact confirmation of discoveries]
\label{rem:exact-confirmation}
    We can strengthen the certificate by confirming each candidate hit with an exact model checking call.
    For requirement $\obj{j}$, recall the true violation event $\unsafeSet{j}$ (\cref{def:signed-gap}) and the inexpensive polynomial candidate event $\vulnRegion{j}$ (\cref{def:conservative-gap}), and define the \emph{confirmed discovery} event to be $\confirmSet{j} = \unsafeSet{j} \cap \vulnRegion{j}$.
    The confirmed non-discovery score after $k$ tests is then
    $\confirmRisk = \sum_{j=1}^{\nobjs} \severity{j} \Indicator[\text{none of the $k$ tests lies in $\confirmSet{j}$}]$.
    Because $\confirmSet{j} \subseteq \vulnRegion{j}$, replacing $\vulnMeasure{j} = \Pr(\vulnRegion{j})$ by any lower bound on $\Pr(\confirmSet{j})$ gives a conservative instance of the stopping rule in \cref{cor:stopping}, and additionally ensures that every counted discovery is a real violation.
    Our experiments in \cref{sec:benchmarks} instantiate this confirmed variant.
\end{remark}

\subsection{Composition of Both Layers}
\label{ssec:e2e}

We now present explicitly the combination of the two layers.
The intuition is straightforward: 
a truly vulnerable parameter valuation can escape the certificate only if either (i)~the scenario approximation missed it (contributing at most $\errorRate_{j}$ in prior mass per requirement) or (ii)~the testing draws failed to hit the conservative region.
The following theorem adds both sources of error: 
the total risk is the risk from not testing enough plus the risk from the surrogate being wrong.

\begin{restatable}[End-to-end non-discovery bound]{theorem}{thmEndToEnd}
\label{thm:endToEnd}
    Under requirement\nbdash{}wise semantics (\cref{sec:introduction}), let each $\approxfun{j}$ be an $(\errorRate_{j}, \significanceLevel_{j}, \margin_{j})$-PAC approximation obtained from an independent scenario LP~\eqref{eq:scenario-lp}, and construct the vulnerability regions $\vulnRegion{j}$ (\cref{def:conservative-gap}) and the non-discovery score $\residualRisk$ (\cref{def:residual-risk}).
    Then, with confidence at least $1 - \sum_{j=1}^{\nobjs} \significanceLevel_{j}$ over the scenario training draws, the following holds simultaneously for all testing budgets $k$:
    \begin{equation}
    \label{eq:e2e}
        \expectation_{\mathrm{test}}[\residualRisk]
        \leq
        \underbrace{\sum_{j=1}^{\nobjs} \severity{j}\,(1-\vulnMeasure{j})^{k}}_{\text{expected testing risk over $\vulnRegion{j}$}}
        +
        \underbrace{\sum_{j=1}^{\nobjs} \severity{j}\,\errorRate_{j}}_{\text{scenario miss mass}}
    \end{equation}
    where $\expectation_{\mathrm{test}}$ denotes expectation over the $k$ i.i.d.\ testing draws from $\priorDist$ only (the scenario samples are held fixed at their realized values).
\end{restatable}

\subsection{Variance and High-Probability Stopping}
\label{ssec:high-probability}

The expectation certificate can be strengthened to a high-probability statement by accounting for the variability of the non-discovery score.
The exact variance depends on pairwise union masses and therefore captures the dependence between requirements whose vulnerability regions overlap.

\begin{restatable}[Exact variance of the non-discovery score]{proposition}{propExactVariance}
    \label{prop:exactVariance}
    For any testing budget $k$, writing $p_{j} = (1-\vulnMeasure{j})^{k}$,
    \[
        \mathrm{Var}_{\mathrm{test}}(\residualRisk)
        = \sum_{j=1}^{\nobjs} \severity{j}^{2}p_{j}(1-p_{j})
         + 2 \sum_{1 \leq i < j \leq \nobjs} \severity{i} \severity{j}
            \Big[(1-Q_{\{i,j\}})^{k} - p_{i} p_{j} \Big],
    \]
    where $Q_{\{i,j\}} = \int_{\vulnRegion{i} \cup \vulnRegion{j}} \priorDist(\evaluation)\,d\evaluation$.
    When the regions are disjoint, $Q_{\{i,j\}}=\vulnMeasure{i}+\vulnMeasure{j}$ and the covariance terms are generally negative, not zero; the displayed formula still applies.
\end{restatable}

The following result avoids estimating pairwise masses by bounding all covariance terms through Cauchy--Schwarz and then applying Cantelli's inequality.

\begin{restatable}[High-probability non-discovery bound via Cantelli's inequality]{corollary}{corHpCertificate}
    \label{cor:hpCertificate}
    For any $\alpha \in (0,1)$ and testing budget $k$, with confidence at least $1 - \sum_{j=1}^{\nobjs} \significanceLevel_{j}$ over the scenario training draws, define
    \begin{align*}
        \mu_{k} &:= \sum_{j=1}^{\nobjs} \severity{j}(1-\vulnMeasure{j})^{k}
            + \sum_{j=1}^{\nobjs} \severity{j} \errorRate_{j},\\
        \bar{V}_{k} &:= \nobjs \sum_{j=1}^{\nobjs} \severity{j}^{2} (1-\vulnMeasure{j})^{k}
        \bigl(1-(1-\vulnMeasure{j})^{k}\bigr).
    \end{align*}
    Then, for any $\targetRisk>\mu_{k}$,
    \[
        \probability_{\mathrm{test}}(\residualRisk \geq \targetRisk)
        \leq \frac{\bar{V}_{k}}{\bar{V}_{k} + (\targetRisk-\mu_{k})^{2}}.
    \]
    Consequently, $\mu_{k} + \sqrt{\bar{V}_{k}(1/\alpha-1)} \leq \targetRisk$ guarantees
    $\probability_{\mathrm{test}}(\residualRisk \leq \targetRisk) \geq 1-\alpha$.
    Moreover, $\bar{V}_{k} \leq \frac{\nobjs}{4} \sum_{j=1}^{\nobjs} \severity{j}^{2}$.
\end{restatable}

Proofs of \cref{prop:exactVariance,cor:hpCertificate} are given in Appendix~\ref{app:missingProofs}.

\subsection{Certification Algorithm}
\label{ssec:algorithm}

\cref{alg:scenario-risk} summarizes the complete certification procedure.
It combines the scenario-approximation layer (\cref{sec:framework}) with the testing layer (\cref{ssec:residual-risk,ssec:e2e}).

\begin{algorithm}[t]
\caption{Scenario-Risk Certification}
\label{alg:scenario-risk}
\begin{algorithmic}[1]
\Require Parametric model $\mathcal{M}$, requirements $\obj{1}, \dots, \obj{\nobjs}$ with thresholds $\tau_{1}, \dots, \tau_{\nobjs}$, prior density $\priorDist$, polynomial degree $d$, PAC parameters $(\errorRate_{j}, \significanceLevel_{j})$, severity weights $\severity{1}, \dots, \severity{\nobjs}$, target risk $\targetRisk$
\Ensure Conservative surrogates $\approxgap{1}, \dots, \approxgap{\nobjs}$, region masses $\vulnMeasure{1}, \dots, \vulnMeasure{\nobjs}$, stopping budget $k^{\star}$
\For{$j = 1, \dots, \nobjs$} \Comment{\textit{scenario approximation, per requirement}}
  \State Compute sample size $\nsamples_{j}$ from \cref{thm:scenario-samples} \Comment{\textit{PAC budget}}
  \State Draw $\nsamples_{j}$ parameter valuations from $\priorDist$ \Comment{\textit{training set}}
  \State Evaluate $\valuefun{j}$ at the sampled valuations via \storm or a \pacpma-compatible scenario backend \Comment{\textit{model checking queries}}
  \State Fit a degree-$d$ surrogate $\approxfun{j}$ and obtain its margin $\margin_{j}$ \Comment{\textit{LP optimum or maximum training residual}}
  \State Construct the conservative signed surrogate $\approxgap{j}$ from \cref{def:conservative-gap} \Comment{\textit{push approximation toward unsafe direction}}
  \State Estimate $\vulnMeasure{j} = \int_{\vulnRegion{j}} \priorDist\vphantom{(\evaluation)\, d\evaluation}$ by numerical integration or Monte Carlo \Comment{\textit{region mass}}
\EndFor
\State Compute $k^{\star}$ using \cref{cor:stopping} \Comment{\textit{testing layer: derive stopping budget from marginal masses}}
\State \Return $\approxgap{1},\dots,\approxgap{\nobjs}$, $\vulnMeasure{1},\dots,\vulnMeasure{\nobjs}$, $k^{\star}$
\end{algorithmic}
\end{algorithm}

\paragraph*{Cost of adding a requirement.}
Each requirement is fitted separately from its own model-checker values.
Adding one requirement therefore adds one regression problem and its model-checker calls; it does not enlarge the polynomial fitted for any existing requirement.
After the region masses have been estimated, the expected non-discovery score is the sum of $M$ terms, one per requirement.
Overlap information is needed only for the exact time to discover every region, not for this marginal-mass score.

\paragraph*{On the practical role of the marginal-mass bound.}
For the profiles evaluated in this paper, we have $M \leq 5$, so the exact formula contains at most $2^{5}-1 = 31$ terms and is inexpensive to evaluate.
The marginal-mass bound $\expectation[\residualRisk] = \sum_{j} \severity{j}(1-\vulnMeasure{j})^{k}$ is nevertheless valuable for three reasons.
First, it provides a \emph{direct stopping rule} that does not require computing the inclusion-exclusion sum or even estimating the pairwise overlaps $Q_{J}$.
Second, it decomposes additively across requirements, making it trivially parallelizable and easy to update incrementally when new requirements are added.
Third, it remains well-defined when $\nobjs$ is large enough to make the exact formula intractable (e.g., in systems with dozens of monitored safety constraints).
In the small-$\nobjs$ regime, the exact formula from \cref{thm:generalizedCoupon} should be preferred;
the marginal-mass bound is the practical fallback for larger instances.

% ============================================================
\section{Experimental Evaluation}
\label{sec:benchmarks}

We evaluate the complete procedure, including scenario approximation, independent probability certification, and random parameter testing. Each test checks whether a fresh parameter valuation lies in each candidate region and uses exact model checking to confirm candidate hits.
We organize the evaluation around the following four research questions:
\begin{enumerate}[label={\bfseries RQ\arabic*.}, ref={RQ\arabic*}, labelindent=0pt, leftmargin=\leftmargini, noitemsep, itemindent=14pt]
\item 
\label{item:rq:trueProbability}
    For fixed candidate regions, how much true-violation probability can lie outside such regions, according to an independent exact certification sample?
\item 
\label{item:rq:covering}
    Do simultaneous upper bounds on the expected non-discovery score cover the values observed in independent repetitions of random parameter testing?
\item 
\label{item:rq:computationCost}
    What is the cost of the complete procedure, and how does it change across increasing model sizes when all experimental settings are held fixed?
\item
\label{item:rq:budgetComparison}
    How do stopping budgets obtained by directly controlling the expected weighted count compare with budgets obtained by allocating the target uniformly across requirements?
\end{enumerate}
Requirement-level results, requirement formulas, and sensitivity analyses are reported in Appendix~\ref{app:moreExperiments}.

\subsection{Benchmarks and Experimental Setup}
\label{ssec:suite}
\label{ssec:setup}
We evaluate our algorithm on ten parametric benchmark families from the \prism benchmark suite~\cite{KNP12b}:
the Bounded Retransmission Protocol (BRP)~\cite{DBLP:conf/papm/DArgenioJJL01}, 
a randomized Consensus protocol~\cite{DBLP:journals/jal/AspnesH90,DBLP:conf/cav/KwiatkowskaNS01}, 
the Crowds anonymity protocol~\cite{DBLP:journals/tissec/ReiterR98,DBLP:journals/jcs/Shmatikov04}, 
the EGL contract-signing protocol~\cite{DBLP:journals/cacm/EvenGL85,DBLP:journals/jcs/NormanS06}, 
the FireWire root-contention protocol~\cite{DBLP:journals/fac/KwiatkowskaNS03}, 
Herman's self-stabilization algorithm~\cite{DBLP:journals/ipl/Herman90}, 
a leader-synchronization protocol (Leader-Sync)~\cite{DBLP:journals/iandc/ItaiR90}, 
NAND multiplexing~\cite{DBLP:journals/tcad/NormanPKS05}, 
the WLAN backoff protocol~\cite{DBLP:conf/papm/KwiatkowskaNS02}, 
and 
the Zeroconf network-configuration protocol~\cite{DBLP:journals/fmsd/KwiatkowskaNPS06,DBLP:journals/rfc/rfc3927}.
The exact requirement formulas and parameter boxes are listed in Appendix~\ref{app:benchmark-properties}.

Our implementation uses \storm~\cite{DBLP:journals/sttt/HenselJKQV22}~1.12.0 and GNU Octave~11.1.0;
all experiments were run on a MacBook Pro with a 10-core Apple M5 processor, 24~GB of memory, and macOS~26.2.
To construct each candidate region, we interact with \storm in batch mode as done in~\cite{DBLP:conf/cav/BadingsJMTJ23,DBLP:conf/atva/LiuTHXZ23} to evaluate every scenario valuation using sound value iteration at absolute precision $10^{-10}$.
Throughout the experimental results, one model checking evaluation means computing one requirement value at one sampled parameter valuation.
The batch interface serves many such evaluations from one initialized model/property session, so the reported evaluation counts are not counts of separate \storm process starts.
For each requirement, we solve the regularized scenario program~\eqref{eq:scenario-qp} over the degree-two polynomial coefficients, with regularization constant $\alpha=10^{-6}$.
We accept the resulting candidate region only after numerically checking every sampled constraint.
Requirement thresholds are frozen constants obtained before the reported runs and are never recomputed when the model size, random seed, or experimental phase changes.
The parameter prior is uniform over the benchmark box: most probability parameters range over $[0.33,0.66]$, while the perturbation parameters of Crowds, Leader-Sync, and WLAN range over $[-0.25,0.25]$.

The experiments instantiate the confirmed-discovery variant of \cref{rem:exact-confirmation}.
Every membership decision for the true violation set $\unsafeSet{j}$ during probability-mass certification and random parameter testing is decided by using the exact rational arithmetic value given by \storm.
The candidate region therefore saves a model-checker call when $\vulnRegion{j}$ is false, while a candidate is counted as a discovery only after exact confirmation of $\unsafeSet{j}$.

The main experiment executes the complete procedure on the first listed configuration of each of the ten families.
Consensus, FireWire, WLAN, and Zeroconf use $M=3$, 4, 4, and 5 requirements, respectively; the other six families use $M=2$, for 28 requirements in total.
The scenario design uses $\errorRate=0.2$ and allocates total significance $0.025$ uniformly across these 28 requirements;
these parameters determine the scenario sample size through \cref{thm:scenario-samples}.
Since our implementation solves the regularized QP~\eqref{eq:scenario-qp} numerically rather than the LP~\eqref{eq:scenario-lp} exactly, we do not claim the PAC guarantee of \cref{thm:scenario-samples} for the fitted regions;
every confidence bound reported below is instead derived from an independent certification sample, as described next.
After solving all scenario programs, we fix the candidate regions and draw 256 new shared parameter valuations for probability-mass certification.
The size 256 is chosen so that, under the Bonferroni allocation below, a requirement with no observed omission still yields a simultaneous upper bound of about $0.03$ on its omitted-violation probability.
For every valuation and every requirement, exact \storm calls determine $U_{j}$, $A_{j}$, and $U_{j}\setminus V_{j}$.
One-sided Clopper--Pearson intervals, Bonferroni-allocated over three quantities for each of 28 requirements, give simultaneous coverage at least $97.5\%$ for the implementation bounds reported below.
Finally, we perform 64 independent repetitions of random parameter testing at $k\in\setnocond{1,2,5,10}$; the parameter valuations in these repetitions are newly drawn and are not taken from the 256 certification valuations.
All severity weights are $1$ and the target expected non-discovery score is $0.1$.

\subsection{Results}
\label{ssec:results}

\begin{table}[t]
\caption{Model sizes and simultaneous certification bounds for fixed candidate regions, together with the observed mean non-discovery score from independent random parameter tests at $k=10$.}
\label{tab:confirmed-main}
\centering
\setlength{\tabcolsep}{5pt}
\begin{tabular}{lrrrrrr}
\toprule
\multicolumn{1}{c}{Benchmark} & \multicolumn{1}{c}{States} & \multicolumn{1}{c}{$\nobjs$} & \multicolumn{1}{c}{$\max m^{\mathrm U}$} & \multicolumn{1}{c}{$\min\underline a^{\mathrm{impl}}$} & \multicolumn{1}{c}{$\overline{\mu}^{\mathrm{cert}}_{10}$} & \multicolumn{1}{c}{$\mu^{\mathrm{obs}}$} \\
\midrule
BRP & 2691 & 2 & 0.031 & 0.257 & 0.077 & 0.016 \\
Consensus & 272 & 3 & 0.031 & 0.300 & 0.064 & 0.000 \\
Crowds & 1145 & 2 & 0.031 & 0.257 & 0.078 & 0.000 \\
EGL & 23863 & 2 & 0.031 & 0.285 & 0.053 & 0.016 \\
FireWire & 4093 & 4 & 0.031 & 0.303 & 0.101 & 0.047 \\
Herman & 9 & 2 & 0.031 & 0.282 & 0.073 & 0.000 \\
Leader-Sync & 26 & 2 & 0.031 & 0.274 & 0.081 & 0.000 \\
NAND & 78332 & 2 & 0.031 & 0.271 & 0.085 & 0.000 \\
WLAN & 3126 & 4 & 0.041 & 0.250 & 0.204 & 0.062 \\
Zeroconf & 1495 & 5 & 0.031 & 0.246 & 0.189 & 0.094 \\
\bottomrule
\end{tabular}
\end{table}

In \cref{tab:confirmed-main}, the column $\max m^{\mathrm U}$ answers \ref{item:rq:trueProbability}: 
it reports, for each benchmark, the largest simultaneous upper bound on the omitted-violation probability $\Pr(U_{j}\setminus V_{j})$ among its requirements.
Twenty-seven of the 28 requirements have no omitted violation in the 256 exact certification valuations and therefore receive the simultaneous upper bound $0.031$; 
one WLAN requirement has one omission, giving $0.041$.
Columns $\overline{\mu}^{\mathrm{cert}}_{10}$ and $\mu^{\mathrm{obs}}$ answer \ref{item:rq:covering}: 
$\overline{\mu}^{\mathrm{cert}}_{10}$ is the certified upper bound on the expected non-discovery score at $k=10$, computed from the per-requirement lower bounds $\underline{a}^{\mathrm{impl}}$ on the confirmed-discovery probability $\Pr(A_{j})$ as $\overline\mu_k^{\mathrm{cert}}=\sum_{j} w_{j}(1-\underline{a}_{j}^{\mathrm{impl}})^{k}$; 
$\mu^{\mathrm{obs}}$ reports the mean of the non-discovery score observed in the independent testing repetitions.
The certified bounds $\overline{\mu}^{\mathrm{cert}}_{10}$ range from $0.053$ to $0.204$, and for every benchmark the observed mean $\mu^{\mathrm{obs}}$ lies below the certified bound.
Because the observed means are computed only from the freshly drawn testing valuations and never from the certification sample, they form an independent empirical check of the certificate.

\begin{table}[tb]
\caption{Cost of the complete procedure by phase; cells report seconds / model-checking evaluations.
The Testing column also shows, in parentheses, the always-check-all count ($64$ repetitions $\times$ $10$ tests $\times$ $M$ requirements).}
\label{tab:confirmed-cost}
\centering
\setlength{\tabcolsep}{5pt}
\begin{tabular}{lrrrrr}
\toprule
\multicolumn{1}{c}{Benchmark} & \multicolumn{1}{c}{$\nobjs$} & \multicolumn{1}{c}{Scenario} & \multicolumn{1}{c}{Certification} & \multicolumn{1}{c}{Testing} & \multicolumn{1}{c}{Total (s)} \\
\midrule
BRP & 2 & 0.7 / 282 & 6.5 / 512 & 1.8 / 137 (1280) & 9.3 \\
Consensus & 3 & 1.6 / 423 & 4.7 / 768 & 1.7 / 257 (1920) & 8.1 \\
Crowds & 2 & 0.6 / 362 & 1.3 / 512 & 0.3 / 131 (1280) & 2.4 \\
EGL & 2 & 2.1 / 282 & 9.7 / 512 & 2.3 / 133 (1280) & 14.3 \\
FireWire & 4 & 1.5 / 564 & 5.5 / 1024 & 1.3 / 253 (2560) & 8.5 \\
Herman & 2 & 0.6 / 362 & 0.1 / 512 & 0.04 / 171 (1280) & 0.8 \\
Leader-Sync & 2 & 0.6 / 362 & 0.1 / 512 & 0.03 / 133 (1280) & 0.8 \\
NAND & 2 & 3.5 / 282 & 75.1 / 512 & 27.7 / 179 (1280) & 106.5 \\
WLAN & 4 & 1.8 / 564 & 9.0 / 1024 & 2.8 / 313 (2560) & 13.7 \\
Zeroconf & 5 & 1.7 / 555 & 19.9 / 1280 & 5.1 / 320 (3200) & 26.8 \\
\bottomrule
\end{tabular}
\end{table}

\Cref{tab:confirmed-cost} uses the same benchmark configurations as \cref{tab:confirmed-main} and reports the time used for scenario-program solution, independent probability-mass certification, and random parameter testing.
Total times range from \qty{0.8}{\s} for Leader-Sync to \qty{106.5}{\s} for NAND.
Exact certification is the largest phase for the two largest-state main configurations, EGL and NAND.
During random parameter testing, checking the candidate predicate first avoids most exact confirmations:
the exact model-checker calls drop to 131--320, compared with the 1,280--3,200 calls that checking every requirement at every test would need ($64$ repetitions $\times$ $10$ tests $\times$ $M$ requirements).
From \cref{tab:confirmed-cost}, we note that certification is slower than scenario: 
this is due to the fact that certification requires a higher number of performed evaluations and the fact that it uses exact rational analysis, which is more expensive than the numerical computation used to construct the scenario approximation, especially for large models.

\begin{figure}[tb]
\centering
    \resizebox{\linewidth}{!}{\includegraphics{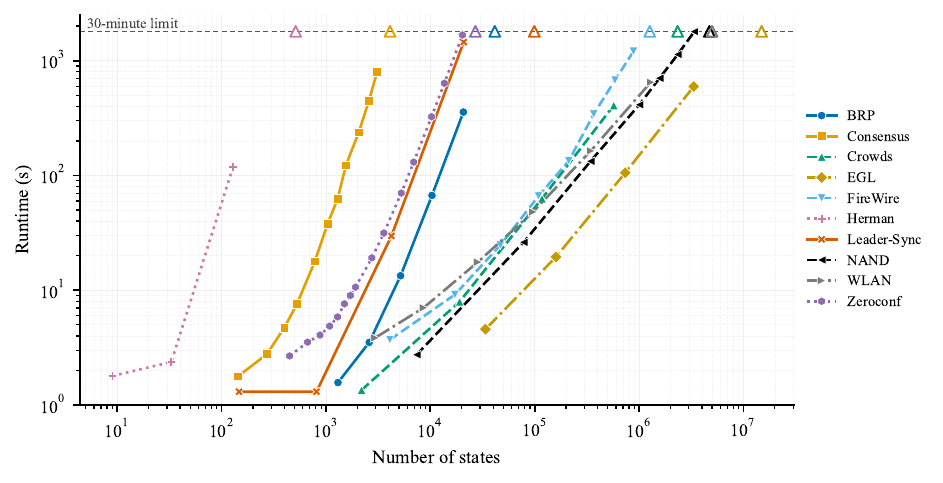}}
    \caption{Runtime versus reachable states for controlled sequences of all ten families; both axes are logarithmic.}
    \label{fig:controlled-runtime-scaling}
\end{figure}

To answer \ref{item:rq:computationCost}, we evaluate all ten benchmark families by fixing the requirements and experimental settings within each family and varying their state space size; 
the outcome is shown in \cref{fig:controlled-runtime-scaling}.
Across the resulting 67 completed executions within 30 minutes, as expected runtime increases with the number of reachable states in every controlled family sequence and grows approximately linearly for the families reaching the largest state spaces.
The completed benchmarks range from 129 states for Herman to more than 3.3 million states for EGL and NAND.
The next attempted instances reached the 30-minute limit in all ten families, represented in \cref{fig:controlled-runtime-scaling} by a open triangle on the dashed timeout line.
These results show that no single state threshold characterizes the implementation limit across all families, although each controlled sequence exhibits the expected increase in runtime as its state space grows.
Interestingly, the first Leader-Sync segment is nearly flat because its 147-state and 812-state instances take almost the same time, where the fixed session-initialization and optimization costs dominate the model checking cost.
For a complementary phase-level view, Appendix~\ref{app:scalingComputationCost} reports illustrative decompositions for Crowds, FireWire, and WLAN: 
total execution time rises from 1.1 to 308.1, 1.7 to 88.0, and 2.2 to 157.7 seconds, respectively, and exact probability-mass certification is the dominant cost at the largest configuration of each sequence.

\subsection{Comparison of Stopping-Budget Rules}
\label{ssec:baseline-detail}

The stopping rule can be compared fairly with another rule only when both use the same probability information and control the same target.
We therefore compare the direct sum rule with uniform per-requirement allocation using the same simultaneous lower bounds $\underline{a}_{j}^{\mathrm{impl}}$ and target $\targetRisk = 0.1$.
Both rules use the same simultaneous lower bounds and control the same expected weighted non-discovery score.
The direct rule constrains the sum of all requirement contributions, whereas uniform allocation reserves an equal share $\targetRisk/M$ for each requirement and may therefore be more conservative.
Their formal definitions and the derivation of the allocation baseline are given in Appendix~\ref{app:uniform-allocation}.

\begin{table}[tb]
\caption{Stopping budgets from the certified lower bounds.}
\label{tab:confirmed-budgets}
\setlength{\tabcolsep}{5pt}
\centering
\begin{tabular}{lrrrr}
\toprule
\multicolumn{1}{c}{Benchmark} & \multicolumn{1}{c}{$\nobjs$} & \multicolumn{1}{c}{$k_{\mathrm{cert}}$} & \multicolumn{1}{c}{$k_{\mathrm{alloc}}$} & \multicolumn{1}{c}{$\widehat{\mathbb{E}[T]}_{\mathrm{emp}}$} \\
\midrule
BRP & 2 & 10 & 11 & 3.92 \\
Consensus & 3 & 9 & 10 & 3.99 \\
Crowds & 2 & 10 & 11 & 3.94 \\
EGL & 2 & 9 & 9 & 3.10 \\
FireWire & 4 & 11 & 11 & 4.13 \\
Herman & 2 & 10 & 10 & 2.61 \\
Leader-Sync & 2 & 10 & 10 & 2.67 \\
NAND & 2 & 10 & 10 & 2.69 \\
WLAN & 4 & 13 & 13 & 4.06 \\
Zeroconf & 5 & 13 & 14 & 4.49 \\
\bottomrule
\end{tabular}
\end{table}

\Cref{tab:confirmed-budgets} answers \ref{item:rq:budgetComparison}.
The direct sum rule (column $k_{\mathrm{cert}}$) never requires more tests than uniform allocation (column $k_{\mathrm{alloc}}$).
It saves one test for BRP, Consensus, Crowds, and Zeroconf and gives the same budget for the remaining six families.
The column $\widehat{\mathbb{E}[T]}_{\mathrm{emp}}$ of \cref{tab:confirmed-budgets} gives the estimate of $\expectation[T]$ obtained by substituting the empirical union masses into \cref{thm:generalizedCoupon}.
Its values, 2.61--4.49 samples, estimate the mean number of draws needed to hit every candidate region under the empirical masses.
They are smaller than the certified budgets of 9--13 samples because the latter use simultaneous lower confidence bounds and control the expected non-discovery score at 0.1, whereas $\widehat{\mathbb{E}[T]}_{\mathrm{emp}}$ is an uncertified mean collection time.
Thus, the values indicate that complete discovery typically occurs after a few draws in the empirical model, but they neither certify $\expectation[T]$ nor select a stopping budget.

\subsection{Scaling Beyond $\nobjs=2$: Synthetic Experiment}
\label{ssec:scaling-detail}

\Cref{tab:confirmed-main} reports benchmark profiles with up to $M=5$ requirements.
To examine larger requirement counts beyond those available in the benchmarks, we use a synthetic common-core overlap model:
each region is $V_{j} = C \cup P_{j}$ where $C$ with $|C| = \rho \cdot \min_{j} q_{j}$ is a shared core and the $P_{j}$'s are disjoint private parts.
Region masses follow a geometric ladder from $q_{\min} = 0.15$ with asymmetry ratio $q_{\max}/q_{\min} = 2$.
We evaluate $\nobjs \in \{2, 4, 8, 16\}$ and overlap fractions $\rho \in \{0, 0.25, 0.5, 0.75\}$, computing the marginal-mass stopping budget $k^{\star}$ from \cref{cor:stopping} and the naive union-bound budget $k_{\text{UB}}$ at $\targetRisk = 0.01$.
Because \cref{cor:stopping} depends only on the marginal masses $q_{j}$, not on $Q_{J}$ for $\abs{J}>1$, the resulting $k^{\star}$ is invariant to $\rho$; evaluating several overlap fractions makes this invariance explicit.

\begin{table}[tb]
\caption{Marginal-mass stopping budget $k^{\star}$ vs.\ naive union-bound $k_{\text{UB}}$ for the synthetic scaling experiment.}
\label{tab:scaling}
\centering
\setlength{\tabcolsep}{5pt}
\begin{tabular}{crrrrrrrrr}
\toprule
\multirow{2}{*}{$\nobjs$} & \multirow{2}{*}{$k_{\text{UB}}$} & \multicolumn{4}{c}{$k^{\star}$} & \multicolumn{4}{c}{Saving vs.\ union bound} \\
\cmidrule(lr){3-6}\cmidrule(lr){7-10}
 & & $\rho{=}0$ & $\rho{=}0.25$ & $\rho{=}0.5$ & $\rho{=}0.75$ & $\rho{=}0$ & $\rho{=}0.25$ & $\rho{=}0.5$ & $\rho{=}0.75$ \\
\midrule
2 & 33 & 29 & 29 & 29 & 29 & 12\% & 12\% & 12\% & 12\% \\
4 & 37 & 30 & 30 & 30 & 30 & 19\% & 19\% & 19\% & 19\% \\
8 & 42 & 33 & 33 & 33 & 33 & 21\% & 21\% & 21\% & 21\% \\
16 & 46 & 36 & 36 & 36 & 36 & 22\% & 22\% & 22\% & 22\% \\
\bottomrule
\end{tabular}
\end{table}

\Cref{tab:scaling} shows that sample savings increase with $\nobjs$, from 12\% at $\nobjs=2$ to 22\% at $\nobjs=16$.
The invariance across $\rho$ is expected: the identical columns confirm that the $\bigO(\nobjs)$ stopping rule depends only on the marginal masses.
The savings therefore come from the geometric asymmetry in region masses and from the summed-expectation stopping criterion, not from the explicit overlap terms.

\begin{table}[tb]
\caption{Asymmetry amplifies savings.}
\label{tab:asymmetry}
\centering
\setlength{\tabcolsep}{5pt}
\begin{tabular}{crrr}
\toprule
$q_{\max}/q_{\min}$ & $k^{\star}$ & $k_{\text{UB}}$ & Saving \\
\midrule
1 & 42 & 42 & 0\% \\
2 & 33 & 42 & 21\% \\
4 & 31 & 42 & 26\% \\
8 & 30 & 42 & 29\% \\
\bottomrule
\end{tabular}
\end{table}

\Cref{tab:asymmetry} isolates the effect of mass asymmetry.
When all requirements have identical mass ($q_{\max}/q_{\min} = 1$), the marginal-mass certificate offers no savings over the union bound.
As asymmetry increases, the certificate exploits the fact that large-mass requirements are effectively ``free'' to discover, reducing the budget by up to 29\% at $q_{\max}/q_{\min} = 8$.

\paragraph*{Computational advantage of $\bigO(\nobjs)$.}
For large $\nobjs$, the exact inclusion-exclusion formula (\cref{thm:generalizedCoupon}) has $\bigO(2^{\nobjs})$ terms, while the marginal-mass certificate (\cref{thm:residualRisk}) remains $\bigO(\nobjs)$.
In our benchmarks, computing the exact formula takes less than \qty{1}{\us} for $\nobjs \leq 4$ but grows to \qty{21}{\ms} at $\nobjs=16$;
the marginal-mass certificate stays under \qty{1}{\us} throughout.
For systems with dozens of concurrent safety constraints, the marginal-mass certificate is the only practical option.

% ============================================================
\section{Conclusion}
\label{sec:conclusion}

In this paper, we presented a framework for multi-property vulnerability discovery under requirement-wise semantics in parametric MDPs.
The central idea is a separation between scenario approximation and testing risk, formalized by the composition theorem (\cref{thm:endToEnd}).
This yields a decomposition of the expected non-discovery score into a testing component that decays geometrically with the number of random parameter tests and a scenario miss-mass component controlled by the approximation layer.
The resulting $\bigO(\nobjs)$ stopping certificate depends only on marginal region masses, while the exact discovery-time and variance formulas account for overlap between vulnerability regions.
The experiments confirm the certificates and the implementation savings for multi-property vulnerability discovery.
% ============================================================

\subsubsection*{Acknowledgements.}
Work supported in part by 
National Key R\&D Program of China No.2025YFE0220300,
the Beijing Natural Science Foundation Project No.\@ IS26039,
the CAS Project for Young Scientists in Basic Research under grant No.\@ YSBR-040,
NSFC under grant No.\@ 61836005,
the CAS Pioneer Hundred Talents Program,
and
the ISCAS New Cultivation Project ISCAS-PYFX-202201.
\newline\protect\includegraphics[height=8pt]{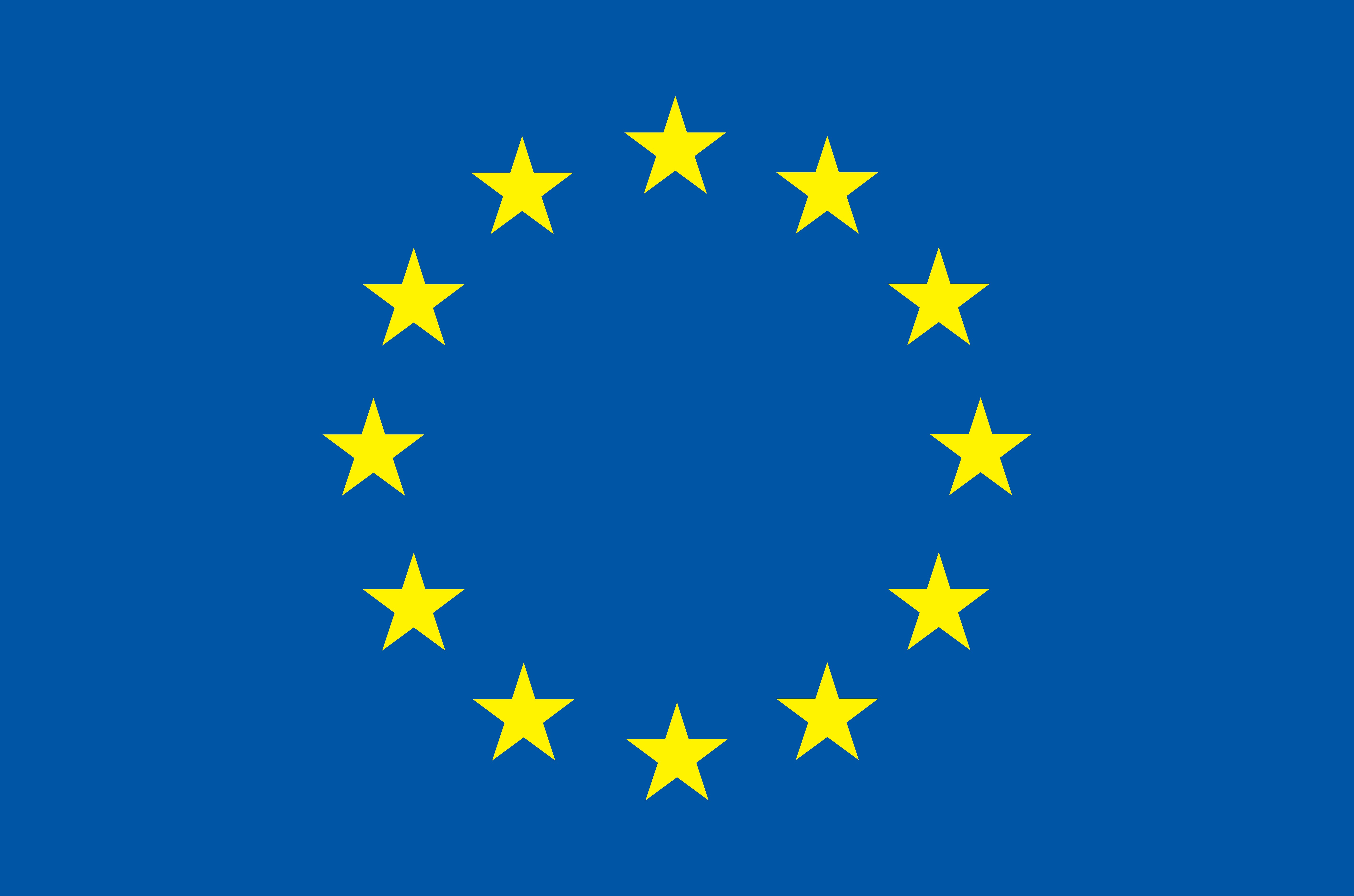} 
This project is part of the European Union's Horizon 2020 research and innovation programme under the Marie Sk\l{}odowska-Curie grant no.\@ 101008233.

\bibliographystyle{splncs04}
\bibliography{refs}

\clearpage
\appendix
\renewcommand*{\theHsection}{appendix.\Alph{section}}
\renewcommand*{\theHsubsection}{\theHsection.\arabic{subsection}}

\section{Missing Proofs}
\label{app:missingProofs}

In this appendix we provide the proofs for the results stated in the main part of the paper.

\propCoverage*
\begin{proof}
    Consider first the upper-bounded case.
    If $\evaluation \in \unsafeSet{j}$, then $\valuefun{j}(\evaluation)>\tau_{j}$.
    Whenever $\abs{\approxfun{j}(\evaluation)-\valuefun{j}(\evaluation)}\leq \margin_{j}$, we obtain
    \[
        \approxfun{j}(\evaluation)+\margin_{j} \geq \valuefun{j}(\evaluation) > \tau_{j},
    \]
    hence $\approxgap{j}(\evaluation)>0$ and therefore $\evaluation \in \vulnRegion{j}$.
    So a violating valuation can be excluded from $\vulnRegion{j}$ only on the PAC bad set where the approximation error exceeds $\margin_{j}$.
    This happens with probability at most $\errorRate_{j}$, with confidence at least $1 - \significanceLevel_{j}$.
    
    The lower-bounded case is symmetric.
\end{proof}

\thmGeneralizedCoupon*
\begin{proof}
    Let $A_{j}^{(k)}$ be the event that region $\vulnRegion{j}$ has not been hit in the first $k$ samples.
    Then $T>k$ if and only if at least one region is still undiscovered, i.e.,
    \[
        \probability(T>k)=\probability\Big(\bigcup_{j=1}^{\nobjs} A_{j}^{(k)}\Big).
    \]
    Applying inclusion-exclusion,
    \[
        \probability(T>k) =
        \sum_{\emptyset \neq J \subseteq \{1,\dots,\nobjs\}}
        (-1)^{|J|+1}
        \probability\Big(\bigcap_{j \in J} A_{j}^{(k)}\Big).
    \]
    The event $\bigcap_{j \in J} A_{j}^{(k)}$ means that none of the first $k$ samples falls in $\bigcup_{j \in J}\vulnRegion{j}$, which has probability $(1-Q_{J})^{k}$ by independence.
    Hence
    \[
        \probability(T>k) =
        \sum_{\emptyset \neq J \subseteq \{1, \dots, \nobjs\}}
        (-1)^{|J|+1} (1-Q_{J})^{k}.
    \]
    Since $T$ is positive integer-valued,
    \[
        \expectation[T]
        = \sum_{t=1}^{\infty} t\probability(T=t)
        = \sum_{t=1}^{\infty} \sum_{k=0}^{t-1} \probability(T=t).
    \]
    Swapping the order of summation yields
    \[
        \expectation[T]
        = \sum_{k=0}^{\infty} \sum_{t=k+1}^{\infty} \probability(T=t)
        = \sum_{k=0}^{\infty} \probability(T>k).
    \]
    Substituting the tail formula and exchanging the finite subset sum with the outer series,
    \[
        \expectation[T]
        = \sum_{\emptyset \neq J \subseteq \{1,\dots,\nobjs\}}
        (-1)^{|J|+1} \sum_{k=0}^{\infty}(1-Q_{J})^{k}
        = \sum_{\emptyset \neq J \subseteq \{1,\dots,\nobjs\}} (-1)^{|J|+1} \frac{1}{Q_{J}},
    \]
    where the last equality uses $Q_{J}>0$ and the geometric-series identity $\sum_{k=0}^{\infty} (1-Q_{J})^{k} = 1/Q_{J}$.
\end{proof}

\thmResidualRisk*
\begin{proof}
    By linearity of expectation,
    \[
        \expectation[\residualRisk] =
        \sum_{j=1}^{\nobjs} \severity{j}
        \probability\big(j \notin \discover{k}\big).
    \]
    The event $j \notin \discover{k}$ means that none of the first $k$ samples lies in $\vulnRegion{j}$.
    Each sample misses $\vulnRegion{j}$ with probability $1-\vulnMeasure{j}$, so independence gives
    \[
        \probability(j \notin \discover{k}) = (1-\vulnMeasure{j})^{k}.
    \]
    Substituting this identity proves the result.
\end{proof}

\thmEndToEnd*
\begin{proof}
    Fix the scenario training samples.
    By \cref{prop:coverage} and the union bound in \cref{cor:simultaneous}, with confidence $1 - \sum_{j} \significanceLevel_{j}$, each true violation set $\unsafeSet{j}$ satisfies
    \[
        \int_{\unsafeSet{j}\setminus\vulnRegion{j}} \priorDist \leq \errorRate_{j}.
    \]
    Now consider a testing draw $\evaluation$.
    If requirement $\obj{j}$ is truly violated ($\evaluation \in \unsafeSet{j}$) but the conservative region does not cover it ($\evaluation \notin \vulnRegion{j}$), the miss is blamed on the scenario error, contributing at most $\errorRate_{j}$ in prior mass.
    Conditionally on the scenario samples satisfying the PAC event, we split the undiscovered vulnerability mass:
    \[
        \probability_{\mathrm{test}}\big(\text{$j \notin \discover{k}$ w.r.t.\@ $\unsafeSet{j}$}]
        \leq
        \probability_{\mathrm{test}}(\text{$j \notin \discover{k}$  w.r.t.\@ $\vulnRegion{j}$})
        + \errorRate_{j}
        = (1-\vulnMeasure{j})^{k} + \errorRate_{j}.
    \]
    Multiplying by $\severity{j}$ and summing over $j$ via linearity of expectation gives~\eqref{eq:e2e}.
\end{proof}

\propExactVariance*
\begin{proof}
Write $I_{j} = \Indicator[j \notin \discover{k}]$.
The sign of the dependence between $I_{j}$ and $I_{j'}$ is determined by the pairwise union mass; overlap decreases that mass and may make the covariance positive.
We have
\[
    \mathrm{Cov}(I_{j}, I_{j'}) = \probability(I_{j}=1, I_{j'}=1) - p_{j} p_{j'}.
\]
The event $\{I_{j}=1, I_{j'}=1\}$ means no testing sample fell in $\vulnRegion{j} \cup \vulnRegion{j'}$, which has probability $(1-Q_{\{j,j'\}})^{k}$ by independence.
Substituting into the variance expansion $\mathrm{Var}(\sum_{j} \severity{j} I_{j}) = \sum_{j,j'} \severity{j}\severity{j'}\,\mathrm{Cov}(I_{j},I_{j'})$ yields the result.
\end{proof}

\corHpCertificate*
\begin{proof}
    We use the one-sided Cantelli (Chebyshev--Cantelli) inequality: 
    for any random variable $X$ with mean $\mu$ and variance $V$,
    \[
        \probability(X - \mu \geq t) \leq \frac{V}{V + t^{2}} \qquad \text{for all $t > 0$}.
    \]
    By \cref{thm:endToEnd}, $\expectation_{\mathrm{test}}[\residualRisk] \leq \mu_{k}$ with the stated scenario confidence.
    We now derive an \emph{upper} bound on $\mathrm{Var}_{\mathrm{test}}(\residualRisk)$.
    Writing $p_{j} = (1-\vulnMeasure{j})^{k}$ and $I_{j} = \Indicator[j \notin \discover{k}]$, we have $\expectation[I_{j}] = p_{j}$ and
    \[
        \mathrm{Var}_{\mathrm{test}}(\residualRisk) = \sum_{j,j'} \severity{j} \severity{j'}\,\mathrm{Cov}(I_{j}, I_{j'}).
    \]
    Since $0 \leq I_{j} \leq 1$, we have $\abs{\mathrm{Cov}(I_{j}, I_{j'})} \leq \sqrt{p_{j} (1-p_{j}) p_{j'} (1-p_{j'})}$ by Cauchy--Schwarz, giving
    \[
        \mathrm{Var}_{\mathrm{test}}(\residualRisk)
        \leq \Big(\sum_{j} \severity{j} \sqrt{p_{j}(1-p_{j})}\Big)^{2}
        \leq \nobjs \cdot \sum_{j} \severity{j}^{2} p_{j} (1-p_{j})
        = \bar{V}_{k}.
    \]
    Now apply Cantelli.
    Since $\expectation[\residualRisk] \leq \mu_{k}$, the deviation $\residualRisk - \expectation[\residualRisk]$ exceeds $\targetRisk - \expectation[\residualRisk] \geq \targetRisk - \mu_{k} > 0$ whenever $\residualRisk \geq \targetRisk$.
    Using the upper bound $\bar{V}_{k}$ on the variance (which enters the numerator and denominator of the Cantelli fraction), we obtain
    \[
        \probability_{\mathrm{test}}(\residualRisk \geq \targetRisk)
    \leq \frac{\bar{V}_{k}}{\bar{V}_{k} + (\targetRisk - \mu_{k})^{2}}.
    \]
    The stopping rule follows by setting $\bar{V}_{k} / (\bar{V}_{k} + (\targetRisk - \mu_{k})^{2}) \leq \alpha$ and solving.
\end{proof}

\section{Derivation of the Uniform-Allocation Baseline}
\label{app:uniform-allocation}

The stopping-budget comparison in \cref{ssec:baseline-detail} evaluates the marginal-mass certificate against a uniform per-requirement allocation baseline.
For a fair comparison, both rules use the same simultaneous lower bounds $\underline{a}_{j}^{\mathrm{impl}}$ and control the same target $\targetRisk$ on the expected weighted non-discovery score.
This appendix formalizes the two rules, derives the allocation budget, and shows why uniform allocation cannot require fewer tests than the direct certificate.

Using the simultaneous lower bounds reported in the experiments, the direct rule is
\[
    k_{\mathrm{cert}}
    = \min\setcond{k \in \naturals}{
    \sum_{j=1}^{\nobjs} w_{j} (1 - \underline{a}_{j}^{\mathrm{impl}})^{k} \leq \targetRisk}
\]
while the uniform-allocation baseline instead selects
\[
    k_{\mathrm{alloc}}
    = \max_{1 \leq j \leq \nobjs}
    \left\lceil
    \frac{\log(\targetRisk/(\nobjs w_{j}))}
    {\log(1 - \underline{a}_{j}^{\mathrm{impl}})}
    \right\rceil.
\]

For requirement $j$, the certified probability of remaining undiscovered after $k$ independent parameter tests is at most $(1 - \underline{a}_{j}^{\mathrm{impl}})^{k}$.
Uniform allocation assigns the weighted contribution of each requirement the allowance $\targetRisk / \nobjs$ and therefore imposes
\[
    w_{j} (1 - \underline{a}_{j}^{\mathrm{impl}})^{k}
    \leq \frac{\targetRisk}{\nobjs}.
\]
For the nontrivial case $0 < \underline{a}_{j}^{\mathrm{impl}} < 1$ and $0 < \frac{\targetRisk}{\nobjs w_{j}} < 1$, which holds for every requirement in \cref{tab:confirmed-budgets}, this inequality is equivalent to
\[
    k \log(1 - \underline{a}_{j}^{\mathrm{impl}})
    \leq \log\frac{\targetRisk}{\nobjs w_{j}},
    \qquad
    k \geq
    \frac{\log\frac{\targetRisk}{\nobjs w_{j}}}
    {\log(1 - \underline{a}_{j}^{\mathrm{impl}})}.
\]
The direction of the second inequality is reversed because $\log(1 - \underline{a}_{j}^{\mathrm{impl}}) < 0$.
Since $k$ is an integer and one common budget must satisfy all $\nobjs$ requirements, rounding each lower bound upward and taking their maximum gives $k_{\mathrm{alloc}}$ in \cref{ssec:baseline-detail}.
Moreover, summing the $M$ allocated inequalities gives
\[
\sum_{j=1}^{M}w_{j}(1-\underline{a}_{j}^{\mathrm{impl}})^{k}
\leq \sum_{j=1}^{M}\frac{\targetRisk}{M}
=\targetRisk.
\]
Thus uniform allocation is sufficient for the same expected-score target, but not necessary: the direct sum rule may distribute the contributions unevenly and therefore satisfies $k_{\mathrm{cert}}\leq k_{\mathrm{alloc}}$.

\section{Additional Experimental Evaluation}
\label{app:moreExperiments}

This appendix gives the exact benchmark requirements, requirement-level confidence bounds, and derived stopping budgets used by the main experiment.
It also reports the fixed-setting scaling study for \ref{item:rq:computationCost} and separate synthetic and sensitivity studies whose protocols are stated locally.

\subsection{Benchmark Requirements and Configurations}
\label{app:benchmark-properties}

The experiments presented in the main part of the paper use as configurations the models BRP, Consensus, Crowds, EGL, FireWire, Herman, Leader-Sync, NAND, WLAN, and Zeroconf, available from the PRISM benchmark suite~\cite{KNP12b}.
Regarding the requirements used in the experiments, we analyzed the properties shown in \cref{tab:properties}.
A ``$\geq$'' in the ``Bound'' column denotes a lower-bounded requirement $\valuefun{j}\geq\tau_{j}$, while a ``$\leq$'' denotes an upper-bounded requirement $\valuefun{j}\leq\tau_{j}$.
Probability parameters use $[0.33,0.66]$ unless the model declares a perturbation parameter: Crowds' \texttt{voe}, all three Leader-Sync parameters, and both WLAN parameters use $[-0.25,0.25]$.
Every box carries the uniform prior.

\begin{table}[tbp]
\caption{The requirements used in the experiments.}
\label{tab:properties}
\centering
\setlength{\tabcolsep}{5pt}
\begin{tabular}{lllc}
\toprule
Benchmark & Requirement & Formula & Bound \\
\midrule
\multirow{2}{*}{BRP} & Pmax no reception & $\lP[\max]{=?}[\lF\;(\texttt{srep}\neq0 \land \neg\texttt{recv})]$ & $\leq$ \\
 & Rmin steps & $\lER[\min]{\texttt{steps}}[\lF\;\texttt{done}]$ & $\leq$ \\
\midrule
\multirow{3}{*}{Consensus} & Pmin heads & $\lP[\min]{=?}[\lF\;\texttt{heads\_consensus}]$ & $\geq$ \\
 & Pmax disagree & $\lP[\max]{=?}[\lF\;\texttt{disagree\_done}]$ & $\leq$ \\
 & Rmin steps & $\lER[\min]{\texttt{steps}}[\lF\;\texttt{finished}]$ & $\leq$ \\
\midrule
\multirow{2}{*}{Crowds} & Pmax observe twice & $\lP[\max]{=?}[\lF\;\texttt{observe\_sender\_twice}]$ & $\leq$ \\
 & Rmin steps & $\lER[\min]{\texttt{steps}}[\lF\;\texttt{complete}]$ & $\leq$ \\
\midrule
\multirow{2}{*}{EGL} & Pmax advantage & $\lP[\max]{=?}[\lF\;\texttt{b\_advantage}]$ & $\leq$ \\
 & Rmin messages & $\lER[\min]{\texttt{messages\_A\_needs}}[\lF\;\texttt{resolved}]$ & $\leq$ \\
\midrule
\multirow{4}{*}{FireWire} & Pmax root & $\lP[\max]{=?}[\lF\;\texttt{node1\_root}]$ & $\geq$ \\
 & Rmin time & $\lER[\min]{\texttt{time}}[\lF\;\texttt{done}]$ & $\leq$ \\
 & Pmin node 2 root & $\lP[\min]{=?}[\lF\;\texttt{node2\_root}]$ & $\geq$ \\
 & Rmin idle time & $\lER[\min]{\texttt{time\_idle}}[\lF\;\texttt{done}]$ & $\leq$ \\
\midrule
\multirow{2}{*}{Herman} & Pmin stable & $\lP[\min]{=?}[\texttt{true}\;\lBU{10}\;\texttt{stable}]$ & $\geq$ \\
 & Rmin steps & $\lER[\min]{\texttt{steps}}[\lF\;\texttt{stable}]$ & $\leq$ \\
\midrule
\multirow{2}{*}{Leader-Sync} & Pmin elected & $\lP[\min]{=?}[\texttt{true}\;\lBU{10}\;\texttt{elected}]$ & $\geq$ \\
 & Rmin rounds & $\lER[\min]{\texttt{num\_rounds}}[\lF\;\texttt{elected}]$ & $\leq$ \\
\midrule
\multirow{2}{*}{NAND} & Pmax bad quality & $\lP[\max]{=?}[\lF\;\texttt{bad\_quality}]$ & $\leq$ \\
 & Rmax value & $\lER[\max]{\texttt{final\_value}}[\lF\;\texttt{done}]$ & $\geq$ \\
\midrule
\multirow{4}{*}{WLAN} & Rmin time & $\lER[\min]{\texttt{time}}[\lF\;\texttt{done}]$ & $\leq$ \\
 & Rmin cost & $\lER[\min]{\texttt{cost}}[\lF\;\texttt{done}]$ & $\leq$ \\
 & Pmax collision limit & $\lP[\max]{=?}[\lF\;\texttt{collision\_limit}]$ & $\leq$ \\
 & Rmin collisions & $\lER[\min]{\texttt{collisions}}[\lF\;\texttt{done}]$ & $\leq$ \\
\midrule
\multirow{5}{*}{Zeroconf} & Pmin fresh & $\lP[\min]{=?}[\lF\;\texttt{fresh\_configured}]$ & $\geq$ \\
 & Rmin time & $\lER[\min]{\texttt{time}}[\lF\;\texttt{configured}]$ & $\leq$ \\
 & Pmax conflict & $\lP[\max]{=?}[\lF\;\texttt{conflict\_configured}]$ & $\leq$ \\
 & Rmin messages & $\lER[\min]{\texttt{messages}}[\lF\;\texttt{configured}]$ & $\leq$ \\
 & Rmin restarts & $\lER[\min]{\texttt{restarts}}[\lF\;\texttt{configured}]$ & $\leq$ \\
\bottomrule
\end{tabular}
\end{table}

\subsection{Confirmed Main-Profile Results}
\label{app:detailed-setup}

\begin{table}[tbp]
\caption{Requirement-level probability estimates and simultaneous confidence bounds.}
\label{tab:confirmed-objectives}
\centering
\setlength{\tabcolsep}{3.7pt}
\begin{tabular}{llrrrrrrr}
\toprule
\multicolumn{1}{c}{Benchmark} & \multicolumn{1}{c}{Requirement} & \multicolumn{1}{c}{$\hat a$} & \multicolumn{1}{c}{$\hat m$} & \multicolumn{1}{c}{$\hat b$} & \multicolumn{1}{c}{$u^L$} & \multicolumn{1}{c}{$a^L$} & \multicolumn{1}{c}{$m^U$} & \multicolumn{1}{c}{$a_{\mathrm{impl}}^L$} \\
\midrule
\multirow{2}{*}{BRP} & Pmax no reception & 0.410 & 0.000 & 0.000 & 0.307 & 0.307 & 0.031 & 0.307 \\
 & Rmin steps & 0.355 & 0.000 & 0.039 & 0.257 & 0.257 & 0.031 & 0.257 \\
\midrule
\multirow{3}{*}{Consensus} & Pmin heads & 0.445 & 0.000 & 0.020 & 0.340 & 0.340 & 0.031 & 0.340 \\
 & Pmax disagree & 0.430 & 0.000 & 0.078 & 0.325 & 0.325 & 0.031 & 0.325 \\
 & Rmin steps & 0.402 & 0.000 & 0.430 & 0.300 & 0.300 & 0.031 & 0.300 \\
\midrule
\multirow{2}{*}{Crowds} & Pmax observe twice & 0.355 & 0.000 & 0.008 & 0.257 & 0.257 & 0.031 & 0.257 \\
 & Rmin steps & 0.406 & 0.000 & 0.023 & 0.303 & 0.303 & 0.031 & 0.303 \\
\midrule
\multirow{2}{*}{EGL} & Pmax advantage & 0.387 & 0.000 & 0.012 & 0.285 & 0.285 & 0.031 & 0.285 \\
 & Rmin messages & 0.434 & 0.000 & 0.004 & 0.329 & 0.329 & 0.031 & 0.329 \\
\midrule
\multirow{4}{*}{FireWire} & Pmax root & 0.418 & 0.000 & 0.000 & 0.314 & 0.314 & 0.031 & 0.314 \\
 & Pmin node 2 root & 0.406 & 0.000 & 0.000 & 0.303 & 0.303 & 0.031 & 0.303 \\
 & Rmin time & 0.410 & 0.000 & 0.000 & 0.307 & 0.307 & 0.031 & 0.307 \\
 & Rmin idle time & 0.410 & 0.000 & 0.000 & 0.307 & 0.307 & 0.031 & 0.307 \\
\midrule
\multirow{2}{*}{Herman} & Pmin stable & 0.383 & 0.000 & 0.324 & 0.282 & 0.282 & 0.031 & 0.282 \\
 & Rmin steps & 0.383 & 0.000 & 0.004 & 0.282 & 0.282 & 0.031 & 0.282 \\
\midrule
\multirow{2}{*}{Leader-Sync} & Pmin elected & 0.375 & 0.000 & 0.012 & 0.274 & 0.274 & 0.031 & 0.274 \\
 & Rmin rounds & 0.375 & 0.000 & 0.004 & 0.274 & 0.274 & 0.031 & 0.274 \\
\midrule
\multirow{2}{*}{NAND} & Pmax bad quality & 0.371 & 0.000 & 0.379 & 0.271 & 0.271 & 0.031 & 0.271 \\
 & Rmax value & 0.371 & 0.000 & 0.031 & 0.271 & 0.271 & 0.031 & 0.271 \\
\midrule
\multirow{4}{*}{WLAN} & Pmax collision limit & 0.387 & 0.004 & 0.168 & 0.289 & 0.285 & 0.041 & 0.285 \\
 & Rmin collisions & 0.348 & 0.000 & 0.188 & 0.250 & 0.250 & 0.031 & 0.250 \\
 & Rmin time & 0.348 & 0.000 & 0.125 & 0.250 & 0.250 & 0.031 & 0.250 \\
 & Rmin cost & 0.348 & 0.000 & 0.125 & 0.250 & 0.250 & 0.031 & 0.250 \\
\midrule
\multirow{5}{*}{Zeroconf} & Pmax conflict & 0.344 & 0.000 & 0.000 & 0.246 & 0.246 & 0.031 & 0.246 \\
 & Pmin fresh & 0.344 & 0.000 & 0.000 & 0.246 & 0.246 & 0.031 & 0.246 \\
 & Rmin time & 0.430 & 0.000 & 0.035 & 0.325 & 0.325 & 0.031 & 0.325 \\
 & Rmin messages & 0.410 & 0.000 & 0.023 & 0.307 & 0.307 & 0.031 & 0.307 \\
 & Rmin restarts & 0.410 & 0.000 & 0.004 & 0.307 & 0.307 & 0.031 & 0.307 \\
\bottomrule
\end{tabular}
\end{table}

\Cref{tab:confirmed-objectives} provides the empirical event frequencies and simultaneous confidence bounds underlying every aggregate row of \cref{tab:confirmed-main}.
Here $\hat a$, $\hat m$, and $\hat b$ estimate $\Pr(U_{j}\cap V_{j})$, $\Pr(U_{j}\setminus V_{j})$, and $\Pr(V_{j}\setminus U_{j})$, respectively.
The values $u^L$ and $a^L$ are direct lower bounds on $\Pr(U_{j})$ and $\Pr(A_{j})$, while $m^U$ is an upper bound on $\Pr(U_{j}\setminus V_{j})$.
The implementation lower bound is $a_{\mathrm{impl}}^L=\max\{a^L,u^L-m^U\}$, and the stopping certificate uses only this quantity.
The largest observed omitted-violation frequency is $1/256$ for WLAN's collision-limit requirement, and every other requirement has zero observed omissions.
Candidate false positives vary much more: for example, the empirical $\hat b$ is $0.430$ for Consensus steps, $0.379$ for NAND quality, and zero for several FireWire and Zeroconf requirements.
This distinction matters operationally: false positives increase exact confirmation calls, whereas omitted violations affect coverage.

\subsection{Scaling Study for Computation Cost}
\label{app:scalingComputationCost}

The separate scaling experiment executes the same three phases, i.e.\@ scenario-program solution, probability-mass certification, and random parameter testing, on three increasing configurations from each of Crowds, FireWire, and WLAN, always with two requirements.
All nine complete executions of this experiment use the same settings: $\errorRate=0.4$, 64 independent certification valuations, 16 independent repetitions of random parameter testing, and $k\in\setnocond{1,2,5}$.
The first-configuration thresholds are reused within a family so that changing model size does not silently change the requirements.
This smaller protocol is used only to measure the cost of complete executions across model sizes; all probability-quality conclusions in the main experiment come from the 28-requirement profile.

\begin{table}[tb]
\caption{Cost of one complete execution in the scaling experiment; cells report seconds / model-checking evaluations.
}
\label{tab:confirmed-scaling}
\centering
\setlength{\tabcolsep}{5pt}
\begin{tabular}{lrrrrr}
\toprule
\multicolumn{1}{c}{Model instance} & \multicolumn{1}{c}{States} & \multicolumn{1}{c}{Scenario} & \multicolumn{1}{c}{Certification} & \multicolumn{1}{c}{Testing} & \multicolumn{1}{c}{Total (s)} \\
\midrule
Crowds (small) & 1145 & 0.6 / 170 & 0.3 / 128 & 0.1 / 56 & 1.1 \\
Crowds (medium) & 42115 & 1.2 / 170 & 15.5 / 128 & 7.0 / 59 & 23.8 \\
Crowds (large) & 321751 & 33.1 / 170 & 184.6 / 128 & 89.9 / 64 & 308.1 \\
FireWire (small) & 4093 & 0.7 / 130 & 0.6 / 128 & 0.3 / 59 & 1.7 \\
FireWire (medium) & 17190 & 1.1 / 130 & 2.6 / 128 & 1.1 / 57 & 4.9 \\
FireWire (large) & 212268 & 11.6 / 130 & 52.6 / 128 & 23.1 / 58 & 88.0 \\
WLAN (small) & 2954 & 0.7 / 130 & 0.9 / 128 & 0.5 / 70 & 2.2 \\
WLAN (medium) & 96302 & 6.2 / 130 & 23.5 / 128 & 14.3 / 78 & 44.2 \\
WLAN (large) & 345000 & 25.5 / 130 & 80.5 / 128 & 50.9 / 82 & 157.7 \\
\bottomrule
\end{tabular}
\end{table}

\Cref{tab:confirmed-scaling} reports the measured cost of one complete execution per configuration at three increasing model sizes for each family.
Within each family the number of scenario and certification calls is fixed by the shared experimental settings, while the state space increases from 1,145 to 321,751 states for Crowds, from 4,093 to 212,268 for FireWire, and from 2,954 to 345,000 for WLAN.
The corresponding complete-execution times increase from 1.1 to 308.1 seconds, from 1.7 to 88.0 seconds, and from 2.2 to 157.7 seconds, respectively.
The phase breakdown shows where this increase occurs: exact probability-mass certification is the dominant cost at the largest configuration of every family, followed by scenario-program solution and exact confirmations during random parameter testing.

To examine this relation over all benchmark families without conflating distinct configurations of the same size, we additionally vary one structural size parameter per family while keeping the other configuration values, the requirement set, and all experimental settings fixed.
The horizontal axis in \cref{fig:controlled-runtime-scaling} reports the resulting number of reachable states, rather than the varied parameter itself.

The figure contains 67 completed executions.
Runtime is non-decreasing with the number of reachable states along every family sequence, which removes the reversals observed when incomparable configurations were ordered only by their state counts.
For the families reaching the largest state spaces, runtime grows approximately linearly with the number of states over the measured range.
The different rates across families also show that state count is not a family-independent predictor of cost; transition structure, requirements, and exact rational computations affect the work per valuation.
This state-space trend is distinct from the $\bigO(\nobjs)$ evaluation cost of the marginal-mass certificate because $\nobjs$, the number of requirements, is fixed within every curve in \cref{fig:controlled-runtime-scaling}.

\begin{table}[tb]
\caption{Exact state count and runtime of the largest completed instance in each controlled family sequence. Every execution has a 30-minute limit.}
\label{tab:max-size-limits}
\centering
\setlength{\tabcolsep}{5pt}
\begin{tabular}{lrr}
\toprule
\multicolumn{1}{c}{Family} & \multicolumn{1}{c}{Largest completed states} & \multicolumn{1}{c}{Time (s)} \\
\midrule
BRP & 20,744 & 356.6 \\
Consensus & 3,088 & 793.0 \\
Crowds & 572,153 & 407.8 \\
EGL & 3,342,334 & 596.1 \\
FireWire & 883,109 & 1203.9 \\
Herman & 129 & 118.8 \\
Leader-Sync & 20,884 & 1453.6 \\
NAND & 3,377,080 & 1789.8 \\
WLAN & 1,295,218 & 648.4 \\
Zeroconf & 20,314 & 1673.2 \\
\bottomrule
\end{tabular}
\end{table}

\Cref{tab:max-size-limits} records the exact largest-completed state count and runtime in each controlled sequence; the main part retains only the trend plot and its summary.

\subsection{Comparison of Candidate-Region Construction Methods}
\label{app:backend-detail}

\begin{table}[tb]
\caption{Numerical maximum-residual construction vs.\@ the certified \pacpma-style construction. The absolute-error and region-mass entries are means over the two requirements of each family; Miss is the maximum.}
\label{tab:backend}
\centering
\setlength{\tabcolsep}{5pt}
\begin{tabular}{l rr r rr rr}
\toprule
\multirow{2}{*}{Benchmark} & \multicolumn{2}{c}{Absolute error} & \multirow{2}{*}{True} & \multicolumn{2}{c}{Conservative} & \multicolumn{2}{c}{Miss} \\
\cmidrule(lr){2-3}\cmidrule(lr){5-6}\cmidrule(lr){7-8}
 & Emp. & PAC. & & Emp. & PAC. & Emp. & PAC. \\
\midrule
BRP & 0.164 & 0.260 & 0.375 & 0.458 & 0.403 & 0.000 & 0.000 \\
Consensus & 0.025 & 0.037 & 0.438 & 0.673 & 0.528 & 0.000 & 0.000 \\
Crowds & 0.010 & 0.019 & 0.395 & 0.423 & 0.398 & 0.000 & 0.000 \\
EGL & 0.005 & 0.007 & 0.413 & 0.433 & 0.415 & 0.000 & 0.000 \\
FireWire & $<$1e-4 & $<$1e-4 & 0.408 & 0.408 & 0.408 & 0.000 & 0.000 \\
Herman & 6e-5 & 1e-4 & 0.470 & 0.798 & 0.585 & 0.000 & 0.000 \\
Leader-Sync & 1e-4 & 2e-4 & 0.390 & 0.463 & 0.423 & 0.000 & 0.000 \\
NAND & 5e-4 & 8e-4 & 0.410 & 0.713 & 0.585 & 0.000 & 0.000 \\
WLAN & 0.356 & 0.468 & 0.400 & 0.605 & 0.540 & 0.000 & 0.000 \\
Zeroconf & $<$1e-4 & $<$1e-4 & 0.393 & 0.428 & 0.408 & 0.000 & 0.000 \\
\bottomrule
\end{tabular}
\end{table}

As a separate sensitivity study, \cref{tab:backend} compares the regularized numerical construction and the \pacpma-style certified construction under identical degree-two settings and on the same 200 independently drawn parameter valuations.
The absolute-error columns report the mean of $|\valuefun{j}(\evaluation)-\approxfun{j}(\evaluation)|$ over the empirical and PAC valuations, respectively; Miss is the measured mass of true violations excluded from the conservative region.
The ``True'' column provides the measured true-violation mass, which is common to both methods because they use the same requirements and parameter valuations; comparing it with each ``Conservative'' value shows the additional mass included by the conservative region.

Both construction methods have zero observed miss mass on these 20 requirements.
The regularized numerical construction has smaller mean absolute error for every requirement and takes \qty{1075}{\s}, compared with \qty{1804}{\s} for the \pacpma-style construction.
The certified construction's mean conservative mass, $0.469$, is closer to the mean true mass, $0.409$, than the regularized numerical construction's $0.540$.
Accordingly, the regularized numerical construction gives a smaller value-function approximation error and shorter runtime, whereas the \pacpma-style construction gives a tighter region and a certified scenario margin in this comparison.

\subsection{Sensitivity to the Polynomial Degree}
\label{app:degree-sensitivity}

Parametric value functions are generally rational or piecewise algebraic rather than degree-two polynomials.
Degree two is therefore an empirical choice that keeps the monomial basis and the scenario sample count smaller than the higher-degree alternatives tested here.

\begin{table}[tb]
\caption{Polynomial-degree sensitivity on the first configuration of four families. Each family cell is the mean absolute value-function error, first averaged over the independently drawn measurement valuations and then over the two requirements; Mean averages the four displayed family values.}
\label{tab:degree-sensitivity}
\centering
\setlength{\tabcolsep}{5pt}
\begin{tabular}{rlrrrrrr}
\toprule
Degree & Scenario $\nsamples$ & Consensus & EGL & Herman & WLAN & Mean & Max miss \\
\midrule
1 & 280--320 & 0.091 & 0.037 & 0.003 & 1.232 & 0.341 & 0.000 \\
2 & 400--560 & 0.024 & 0.006 & 6e-05 & 0.349 & 0.095 & 0.000 \\
3 & 560--960 & 0.022 & 5e-04 & 6e-05 & 0.344 & 0.091 & 0.000 \\
4 & 760--1560 & 0.009 & 5e-05 & 2e-06 & 0.175 & 0.046 & 0.000 \\
5 & 1000--2400 & 0.009 & 7e-06 & 2e-06 & 0.185 & 0.048 & 0.000 \\
\bottomrule
\end{tabular}
\end{table}

The degree sweep uses the same four model configurations, parameter boxes, PAC parameters, 1,000 independently drawn parameter valuations for measurement, and regularized numerical construction at every degree.
Smaller mean absolute value-function error and miss mass are preferable.
As shown in \cref{tab:degree-sensitivity}, every degree has zero observed miss.
Moving from degree one to two reduces the mean absolute value-function error from $0.341$ to $0.095$.
Degree three reduces it only to $0.091$, while degrees four and five reach $0.046$ and $0.048$.
This additional accuracy raises the maximum per-family number of scenario valuations from 560 at degree two to 1,560 at degree four and 2,400 at degree five.
Degree two therefore captures most of the improvement over a linear polynomial while retaining a substantially smaller basis and scenario-sample count.

\end{document}